\documentclass[twocolumn,10pt]{revtex4}
\usepackage{amsmath,amssymb,amsfonts}
\usepackage{subfigure,multirow}
\usepackage{graphicx}

\def\dd{\mathrm d} 
\def\yk{\mathbf Y_k}
\def\ykt{{\mathbf Y_k}^T} 
 
\def\bN{\mathbf N} 
\def\bR{\mathbf R} 
\def\Sset{\mathbb S} 

\def\Lset{\mathbb L} 
 
\def\Rset{\mathbb R}
\def\mcN{\mathcal N} 
\def\mcM{\mathcal M} 
\def\mcW{\mathcal W}
 
\def\like{\mathcal L} 
\def\lm{_{\ell m}} 
\def\tr{\text{tr }}
\def\ie{\emph{i.e. }} 
\def\eg{\emph{e.g. }}
\def\esp{\mathbb E} 
\def\var{\mathrm{var}} 
\def\cov{\mathrm{cov}} 
\def\lmin{\ell_{\min}}
\def\lmax{\ell_{\max}} 
\def\npix{N_{\text{pix}}}

\def\ji{^{(j)}}
\def\npixj{N_{\text{pix}}\ji}
\def\nside{\texttt{nside}}
\def\fsky{\ensuremath{f_{\text{sky}}}}

\def\varcosm{V_{\text{cosmic}}}
\def\varsample{V_{\text{sample}}}
\def\iid{\stackrel{\textrm{\tiny{i.i.d}}}{\sim}}
\def\1{{\mathbf{1}}} 
\def\ind{{\mathbf{1}}} 
{}{}
\newtheorem{prop}{Proposition}{\bf}{\it}
\newtheorem{rem}{Remark}{\bf}{\it}
\usepackage{natbib}

\newcommand\nmc{400}
\newcommand\nmcSF{100}

\begin{document}

\title{CMB power spectrum estimation using wavelets}

\author{G. Fa\"y}
\altaffiliation{Laboratoire Paul Painlev\'e, UMR 8524,\\ Universit\'e
  Lille 1 and CNRS, 59 655 Villeneuve d'Ascq Cedex, France}
  \email{gilles.fay@univ-lille1.fr} 
\author{F. Guilloux}
\altaffiliation{MODAL'X,
  Universit\'e Paris Ouest -- Nanterre La D\'efense, 200 avenue de
  la R\'epublique, 92001 Nanterre Cedex, France and Laboratoire de Probabilit\'es et
  Mod\`eles Al\'eatoires, UMR 7599, Universit\'e Paris 7 - Denis Diderot and
  CNRS, 175 rue du Chevaleret, 75013 Paris, France}
\author{M. Betoule}

\author{J.-F. Cardoso}
\altaffiliation{ Laboratoire de
  Traitement et Communication de l'Information , UMR 5141, T\'el\'ecom
  ParisTech and CNRS, 46 rue Barrault, 75634 Paris Cedex, France }

\author{J. Delabrouille}

\author{M. Le Jeune}
\affiliation{Laboratoire AstroParticule et Cosmologie, UMR 7164,\\ Universit\'e
   Paris 7 - Denis Diderot and CNRS,\\ 10, rue A. Domon et L. Duquet, 75205 Paris
   Cedex 13, France }

 \begin{abstract}
Observations of the Cosmic Microwave Background (CMB) provide increasingly accurate information about the structure of the Universe at the recombination epoch.  Most of this information is encoded in the angular power spectrum of the CMB. The aim of this work is to propose a versatile and powerful method for spectral estimation on the sphere which can easily deal with non-stationarity, foregrounds and multiple experiments with various specifications. In this paper, we use needlets (wavelets) on the sphere to construct natural and efficient spectral estimators for partially observed and beamed CMB with non stationary noise.  In the case of a single experiment, we compare this method with Pseudo-$C_\ell$ methods. The performance of the needlet spectral estimators (NSE) compares very favorably to the best Pseudo--$C_\ell$ estimators, over the whole multipole range.  On simulations with a simple model (CMB + uncorrelated noise with known variance per pixel + mask), they perform uniformly better.  Their    distinctive ability to aggregate many different experiments, to control the propagation of errors and to produce a single wide-band error bars is highlighted. The needlet spectral estimator is a powerful, tunable tool which is very well suited to angular power spectrum estimation of spherical data such as incomplete and noisy CMB maps.
 \end{abstract}

\date{\today} 
\maketitle

\section*{Introduction}

The estimation of the temperature and polarization angular power
spectra of the Cosmic Microwave Background (CMB) is a key step for
estimating the cosmological parameters.  Cosmological information is
encoded in the huge data sets (time order scanning data or high
resolution maps) provided by ground-based, balloon-borne or satellite
experiments. 

In the ideal case of noiseless and full sky experiments, angular power
spectrum estimation is a straightforward task. The empirical spectrum
of the outcome of a Gaussian stationary field $X$, given by
\begin{equation}
  \label{eq:3}
  \widehat C_\ell = \frac1{2\ell+1} \sum_{m=-\ell}^{\ell} \langle X , Y_{\ell m}
  \rangle ^2 ,
\end{equation}
where $(Y_{\ell m})$ denote the usual spherical harmonics, also is the
\emph{maximum likelihood} estimator of the power spectrum of $X$.  It is
efficient in the sense that its variance reaches the Cram\'er-Rao lower bound.

CMB maps are however more or less strongly contaminated by
foregrounds and instrumental noises, depending on the wavelength,
angular frequency $\ell$ and the direction of observation.
Ground-based experiments cover small parts of the sky while space
missions (COBE, W-MAP and, in the near future, Planck) provide full sky
maps of the CMB, but still contaminated with galactic residuals.
Then, the plain estimate~(\ref{eq:3}) is no longer efficient nor even
unbiased.  
To circumvent the non-stationarity of actual observations, the main
ingredients for the spectral estimation used by, for instance, the
W-MAP collaboration~\cite[7]{Hinshaw+2006} and also in most other
analysis, are broadly the following ones.
Usually, some part of the covered sky is blanked to remove the most
emissive foregrounds or the most noisy measurements.  This amounts to
applying a \emph{mask} or more generally a weight function to the sky.
Most of the emissive foregrounds can be subtracted using some
component separation procedure (see \eg \cite{leach_etal:2008} for
comparison methods with Planck-like simulated data).  Even the best
foreground-subtracting maps require masking a small fraction of the
sky. 
Missing or masked data makes the optimal estimation of the power
spectrum a much harder task.  In particular, it breaks the diagonal
structure of the covariance of the multipole moments $a\lm : = \langle X , Y\lm \rangle$
of any stationary component.  \emph{Maximum-likelihood} estimation of
the spectrum in the pixel domain has a numerical complexity that
scales as $\npix^3$ and requires the storage of a $\npix^2$ matrices.
This is untractable for high resolution experiments such as W-MAP or
Planck ($\npix \simeq 13.10^6$).  Nevertheless, for very low $\ell$'s
($\ell\leq 30$), ML estimation in the pixel domain can be performed on
downgraded resolution maps;
see~\cite{tegmark:1997,bond:jaffe:knox:1998}, for instance.  
At higher $\ell$'s, a sub-optimal method based on the
\emph{Pseudo-$C_\ell$} (PCL) gives quite satisfactory results in terms
of complexity and accuracy \cite{hivon:etal:2002}.  It debiases the
empirical or (pseudo) spectrum from the noise contribution and
deconvolves it from the average mask effect.  It works in the
spherical harmonic domain, uses fast spherical harmonic (SH)
transforms and scales as $\npix^{3/2}$.  The available pixels can be
weighted according to the signal to noise ratio (SNR) at any given
point.
For signal-dominated frequencies (low $\ell$'s), the data are
uniformly weighted; it yields the Pseudo-$C_\ell$ estimator with
uniform weights (PCLU).  At noise-dominated frequencies (high
$\ell$'s), each pixel is weighted by the inverse of the variance of
the noise (PCLW estimator).
The W-MAP collaboration used uniform weights for $\ell\leq 500$, the
inverse of the noise variance for $\ell>500$ (see~\cite[Section
7.5]{Hinshaw+2006}) for its three-year release.
Efstathiou (2004)\nocite{efstathiou:2004} showed that the PCLW
estimator is statistically equivalent to the ML estimator in the low
SNR limit, which is usually the case at high $\ell$'s.  In the same
paper he proposed an hybrid method with a smooth transition between
the two PCL regimes.
Finally, when several maps are available, it is worth considering
\emph{cross-power spectra} between different channels since noise is
usually uncorrelated from channel to
channel~(see~\cite[A1.1]{Hinshaw+2003} or
\cite{polenta:etal:2005}).

Other estimation procedures do not fit in any of the two categories
above.  Among them, the spectral estimation from time ordered data by
\cite{wandelt:larson:lakshminarayanan:2004} or Gibbs sampling and
Monte Carlo Markov chain methods such as MAGIC or Commander, see
\cite{Eriksen:2004ss}.  Those last methods try to estimate the
complete posterior joint probability distribution of the power
spectrum through sampling, which in turn can provide point estimates
of the spectrum but also covariance estimates, etc.  Recently, the
multi-taper approach has been imported from the time series literature
to the field of spherical data by \cite{Wieczorek+2007, Dahlen+2008}.
The goal of this approach is to provide an estimation of a localized
power spectrum, in a noiseless experiment.

In this paper, we focus on spectral estimation of the global power
spectrum, in a frequentist framework.  We consider spectral estimation
at small angular scales, \ie in the range of multipoles where the cost
of ML estimation is prohibitive.  We compare our method to PCL
methods.  We adopt somehow realistic models that include partial
coverage of the sky, symmetric beam convolution, inhomogeneous and
uncorrelated additive pixel noise and multiple experiments.

Localized analysis functions such as wavelets are natural tools to
tackle non-stationarity and missing data issues.
There are different ways to define wavelets (in the broad sense of
space-frequency objects) on the sphere, and our choice is to use the needlets,
the statistical properties of which have received a recent rigorous treatment
(\cite{baldi_etal:2007,baldi:etal:2008a,baldi:etal:2008b}) and which have
already been applied successfully to cosmology (\cite{marinucci_etal:2008,
  pietrobon:balbi:marinucci:2006,delabrouille:etal:2008}).

Needlets benefit from perfect (and freely adjustable) localization in the
spherical harmonic domain, which enables their use for spectral estimation.
Moreover, the correlation between needlet coefficients centered on two fixed
directions of the sky vanishes as the scale goes to infinity, \ie as the
needlet concentrate around those points. The spatial localization is
excellent. This property leads to several convergence results and motivates
procedures based on the approximation of decorrelation between coefficients. In
this contribution, we define and study a new angular power spectrum estimator
that uses the property of localization of the wavelets in both spatial and
frequency domains.

In the case of a single experiment with partial coverage and
inhomogeneous noise, the needlet-based estimator deals straightforwardly
with the variation of noise level over the sky, taking advantage of
their localization in the pixel domain.
Moreover, it allows  a joint spectral estimation from multiple experiments with
different coverages, different beams and different noise levels.
The proposed method  mixes observations from all experiments with spatially varying
weights to take into account the local noise levels.  The resulting spectral
estimator somehow mimics the maximum likelihood estimator based on all the
experiments.

The paper is organized as follows.  
In Section~\ref{sec:framework}, we present the observation model and
recall the basics of needlet analysis and the properties of the
needlet coefficients which are the most relevant for spectral
estimation.  
In Section~\ref{sec:needl-spectr-estim}, we define the needlet
spectral estimators (NSE) in the single-map and multiple-maps
frameworks.
In Section~\ref{sec:monte-carlo-studies}, we present results of Monte
Carlo experiments which demonstrate the effectiveness of our approach.
In Section~\ref{sec:discussion}, we summarize the strong and weak
points of our method and outline the remaining difficulties.

\section{Framework}
\label{sec:framework}

\subsection{Observation model}
\label{sec:observation-model}

Let $T$ denote the temperature anisotropy of the CMB emission. For the
sake of simplicity, we consider the following observation model
\begin{multline}
  \label{eq:model}
  X(\xi_k) = W(\xi_k) \left( (B * T)(\xi_k) + \sigma(\xi_k) Z_k \right)   ,
\\   \; k = 1,\dots,\npix  
\end{multline}
where $(\xi_k)$ is a collection of pixels on the sphere, $W$ denotes a
(0-1)-mask or any weight function $0\leq W \leq 1$, $B$ denotes the
instrumental beam.
An additive instrumental noise is modelled by the term
$\sigma(\xi_k)Z_k$  with the assumption that $(Z_k)$ is an
independent standard Gaussian sequence.  Further, we assume that
$\sigma$, $W$ and $B$ are known deterministic functions and that $B$
is axisymmetric. 
Typically, the variance map $\sigma^2$ writes $\sigma^2(\xi_k) =
\sigma_0^2/N_{\text{obs}}(\xi_k)$, where $N_{\text{obs}}$ is referred
to as the \emph{hit map}, that is, $N_{\text{obs}}(\xi_k)$ is the
number of times a pixel in direction $\xi_k$ is seen by the
instrument.  We assume that the observations have been cleaned from
foreground emissions or that those emissions are present but
negligible outside the masked region.
When observations from several experiments are jointly considered, the
model becomes
\begin{multline}
  \label{eq:model2}
  X_e(\xi_k) = W_e(\xi_k) \left( B_e * T(\xi_k) + \sigma_{e}(\xi_k)Z_{k,e}) \right) 
  \\ 
    k = 1,\dots,\npix \; , \;
    e = 1,\dots, E
\end{multline}
where $e$ indexes the experiment.  The CMB sky temperature $T$ is the
same for all experiments but the instrumental characteristics (beam,
coverage) differ (see for example Table~\ref{tab:fusionparams}), and
the respective noises can usually be considered as independent.

\subsection{Definition and implementation of a needlet analysis}

We recall here the construction and practical computation of the
needlet coefficients.  Details can be found in Guilloux {\it et al.}
(2007\nocite{guilloux:fay:cardoso:2008}; see also Narcowich {\it et
  al.}, 2006\nocite{narcowich:petrushev:ward:2006}).

Needlets are based in a decomposition of the spectral domain in bands
or `scales' which are traditionally index by an integer $j$.
Let $b_\ell\ji$ be a collection of window functions in the multipole
domain, with maximal frequencies $\ell\ji_{\max}$ (see
Figure~\ref{fig:windowfunctions} below).  Consider some
pixelization points $\xi\ji_k, k=1,\dots,\npixj$, associated with
positive weights $\lambda\ji_k, k = 1,\dots,\npixj$ which enable exact
discrete integration (quadrature) for spherical harmonics up to degree
$2\lmax\ji$, that is, equality
\begin{equation*}
  \int_{\Sset} Y\lm(\xi) \dd \xi
  = \sum_{k=1}^{\npixj} \lambda_k\ji Y\lm(\xi\ji_k)
\end{equation*}
holds for any $\ell, m$ such that $\ell \leq 2\ell\ji_{\max}$, $|m|
\leq \ell$.
Needlets are the axisymmetric functions defined by
\begin{equation}
  \label{eq:defneedlet}
  \psi\ji_k(\xi)=\sqrt{\lambda\ji_{k}}\sum_{\ell=0}^{\lmax\ji}
  b\ji_\ell L_\ell (\xi \cdot \xi\ji_k ),
\end{equation}
where $L_\ell$ denote the Legendre polynomial of order $\ell$
normalized according to the condition
$L_\ell(1)=\frac{2\ell+1}{4\pi}$.
For proper choices of window functions $\{b_\ell\ji\}_j$, the family
$\left\{\psi_k\ji\right\}_{k,j}$ is a \emph{frame} on the Hilbert space of
square-integrable functions on the sphere $\Sset$. 
In a $B$-adic scheme, it is even a \emph{tight frame}
\citep{narcowich:petrushev:ward:2006}. Though redundant, tight frames
are complete sets which have many properties reminiscent of
orthonormal bases (see \eg \cite{daubechies:1992}, chap.3).

For any field $X$ on the sphere, the coefficients $\gamma\ji_k :=
(\lambda_k\ji)^{-1/2}\langle X , \psi\ji_k \rangle$ are easily
computed in the spherical harmonic domain as made explicit by the
following diagram
\begin{equation}
  \label{eq:diagram}
  \begin{array}{rcc}
    \left\{X(\xi_k)\right\}_{k=1,\dots,\npix} & \stackrel{\textrm{SHT}}{\longrightarrow} &
    a\lm
    \\
    & & \Downarrow \times
    \\
    \left\{\gamma_k\ji)\right\}_{k=1,\dots,\npixj} &
    \stackrel{\; \; \textrm{SHT}^{-1}}{\Longleftarrow} & b\ji_\ell a\lm
  \end{array}
\end{equation}
Double arrows denotes as many operations (\textit{e.g.} spherical transforms)
as bands.  The initial resolution must be fine enough to allow an exact
computation of the Spherical Harmonics Transform up to degree $\lmax\ji$. If,
say, the HEALPix pixelization is used, the $\nside{}$ parameter of the original
map determines the highest available multipole moments and, in turn, the
highest available band $j$.

\subsection{Distribution of the needlet coefficients}

A square-integrable random process $X$ on the sphere is said to be centered and
stationary (or isotropic) if $\esp(X(\xi)) = 0$, $\esp(X(\xi)^2)<\infty$ and
$\esp(X(\xi)X(\xi')) = (4\pi)^{-1} \sum_\ell C_\ell L_\ell(\xi \cdot \xi')$,
with $C_\ell$ referred to as the angular power spectrum of  $X$.
The next proposition summarizes the first and second order statistical
properties of the needlet coefficients of such a process.  They are the
building blocks for any subsequent spectral analysis using needlets.
\begin{prop}
  \label{prop:coeff-need-gauss}
  Suppose that $X$ is a stationary and centered random field with power spectrum
  $C_\ell$.  Then the needlet coefficients are centered random variables and, for
  any 4-tuple $(j,j',k,k')$
  \begin{gather}
    \cov[\gamma_k\ji, \gamma_{k'}^{(j')}] =
    \sum_{\ell\geq0} b_\ell\ji
    b_\ell^{(j')} C_\ell  L_\ell\left(\cos\theta\right)\label{eq:covgammajk}
  \end{gather}
  where $\theta=\theta(j,k,j',k')$ is the angular distance between
  $\xi_k\ji$ and $\xi_{k'}^{(j')}$.
  In particular
  \begin{equation}
    \label{eq:varneedcoeff}
    \var[\gamma_k\ji] = C\ji \ .
  \end{equation}
  where 
  \begin{equation}
    \label{eq:defcj}
    C\ji := (4\pi)^{-1}\sum_{\ell\geq0}
    \left(b_\ell\ji\right)^2 (2\ell+1) C_\ell.  
  \end{equation}
\end{prop}
In other words, the variance of the coefficients $\gamma_k\ji$ is the
power spectrum of $X$ properly integrated over the $j$-th band.
\begin{rem}
  It also follows from~(\ref{eq:covgammajk}) that if the bands $j$ and
  $j'$ are non-overlapping (this is the case for any non-consecutive
  filters of Figure~\ref{fig:windowfunctions}), all the pairs of
  needlet coefficients $\gamma_k\ji$ and $\gamma_{k'}^{(j')}$ are
  uncorrelated and then independent if the field is moreover Gaussian.
\end{rem}

Suppose now that 
\begin{equation*}
  X(\xi_k)=\sigma(\xi_k)Z_k,\ k=1,\cdots, \npix,
\end{equation*}
is a collection of independent random variables with zero mean and
variance $\sigma^2(\xi_k)$, where $\sigma$ is a band-limited
function. This is a convenient and widely used model for residual
instrumental noise (uncorrelated, but non-stationary).
Needlet coefficients are centered and, moreover, if the quadrature
weights are approximately uniform ($\lambda_k \simeq
4\pi/N_{\textrm{pix}}$, as is the case of HEALPix) and $\sigma$ is
sufficiently smooth, then
\begin{equation*}
  \cov[\gamma_k\ji,\gamma_{k'}^{(j')}]
  \simeq
  \int_\Sset
  \sigma^2(\xi)\psi_k\ji(\xi)\psi_{k'}^{(j')}(\xi)\dd\xi.
\end{equation*}
We denote 
\begin{equation}
  n_k\ji(\sigma)
  :=
  \left(
    \int_\Sset\sigma^2(\xi)|\psi_k\ji(\xi)|^2\dd\xi
  \right)^{1/2}.
\end{equation}
the standard deviation of the needlet coefficient of scale $j$ centered on
$\xi_k$.
When the noise is homogeneous ($\sigma$ is constant), it reduces to
$\frac{\sigma^2}{\npix}\sum_{\ell\geq0}
(2\ell+1)\left(b_\ell\ji\right)^2$.

\subsection{Mask and beam effects}\label{sec:appr-deal-with}

As already noticed in the Introduction, missing or masked data makes
the angular power spectrum estimation a non trivial task. Simple
operations in Fourier space such as debeaming become tricky.  Needlets
are also affected by both the mask and the beam.  The effect on
needlets of beam and mask can be approximated as described below.
These approximations, which lead to simple implementations, are
validated in numerical simulations in relatively realistic conditions
in Section~\ref{sec:monte-carlo-studies}.

\paragraph{Mask.}  Recalling that the needlets are spatially
localized, the needlet coefficients are expected to be insensitive to
the application of a mask on the data if they are computed far away
from its edges.  Numerical and theoretical studies of this property
can be found in \cite{baldi:etal:2008a} and
\cite{guilloux:fay:cardoso:2008}.  In practice, we choose to quantify
the effect of the mask on a single coefficient $\gamma_k\ji$ by the
loss induced on the $\mathbb L^2$-norm of the needlet $\psi_{k}\ji$,
\ie a purely geometrical criterion.  More specifically, needlet
coefficients at scale $j$ are deemed reliable (at level $t\ji$) if
they belong to the set
\begin{equation}
  \label{eq:Wt}
  \mathcal{K}\ji_{t\ji}
  :=
  \left\{
    k=1,\dots,\npixj\; :\; \frac{\|W\psi\ji_k\|^2_2}{\|\psi\ji_k\|^2_2} \geq t\ji
  \right\}
\end{equation}
Parameter $t\ji$ is typically set to $0.99$ or $0.95$ for all bands.  Note that
$t\ji \mapsto \mathcal{K}\ji_t$ is decreasing, $\mathcal{K}\ji_0 = K\ji$ and
$\mathcal{K}\ji_{1^+} = \emptyset$. 
In practice, this set is computed by thresholding the map obtained by the
convolution of the mask with the axisymmetric kernel $\xi \mapsto
\left(\sum_\ell b_\ell\ji L_\ell(\cos \theta)\right)^2$. This operation is easy
to implement in the multipole domain.

\paragraph{Beam.}  Consider now the effect of the instrumental beam.
Its transfer function $B_\ell$ is assumed smooth enough that it can be
approximated in the band $j$ by its mean value $B\ji$ in this band,
defined by
\begin{equation}
  \label{eq:defBj}
  (B\ji)^2
  := 
  (4\pi)^{-1}\sum_{\ell \geq 0} (2\ell + 1) (b_\ell\ji)^2  B_\ell^2 
  .
\end{equation}
In the following, the beam effect for spectral estimation is taken
into account in each band.  Indeed, with Definition~(\ref{eq:defBj}),
with no noise, no mask and a smooth beam, Eq.~(\ref{eq:varneedcoeff})
translates to
\begin{equation}
  \label{eq:varneedcoeffdebeam}
   \var\left[ \frac{\gamma_k\ji}{B\ji}\right] \simeq  C\ji \; , \;
  k=1,\dots,\npixj.
\end{equation}
In other words, thanks to the relative narrowness of the bands and to
the smoothness of the beam and CMB spectrum, the attenuation induced
by the beam can be approximated as acting uniformly in each band and
not on individual multipoles. Numerically, with typical beam values from WMAP
or ACBAR experiments (see Table~\ref{tab:fusionparams}), the relative
difference (statistical bias) between the goal quantity $C\ji$ and the
estimated one $\var(\gamma_k\ji)/(B\ji)^2$ remains under 1\% for bands below
$j=27$ ($\lmax = 875$) for WMAP-W, and below $j=39, \lmax=2000$ for ACBAR.

\section{The needlet spectral estimators (NSE)}
\label{sec:needl-spectr-estim}

\subsection{Smooth spectral estimates from a single map}
\label{sec:spectr-estim-single}

For any sequence of weights $w_k\ji$ such that $\sum_{k=1}^{\npixj}
w_k\ji = 1$ and for a clean (contamination-free), complete (full-sky)
and non-convolved (beam-free) observation of the CMB, the quantity
\begin{equation*}
  \widehat C\ji 
  :=
  \sum_{k=1}^{\npixj} {w}_k\ji \left(\gamma_k\ji\right)^2
\end{equation*}
is an unbiased estimate of $C\ji$, a direct consequence of
Proposition~\ref{prop:coeff-need-gauss}.
\begin{rem}
  For uniform weights, this estimator is nothing but the estimator
  $\widehat C_\ell$ from Eq.~(\ref{eq:3}) binned by the window
  function $(b_\ell\ji)^2$.  
  Indeed, (see diagram~(\ref{eq:diagram}))
  \begin{align}
    \widehat C\ji =&  (4\pi)^{-1} \sum_{\ell \geq 0}
    (b_\ell\ji)^2\sum_{m=-\ell}^{\ell} a_{\ell m}^2\\
    =&   (4\pi)^{-1} \sum_{\ell \geq 0}
    (b_\ell\ji)^2(2\ell+1) \widehat C_\ell
  \end{align}
  This is the uniformly minimum variance unbiased estimator of $C\ji$.
  The so-called \emph{cosmic variance} is the Cram\'er-Rao lower bound
  for estimation of the parameter $C_\ell$ in the full-sky, noise-free
  case. Its expression simply is $2C_\ell^2/(2\ell+1)$.  Its
  counterpart for the binned estimator $C\ji$ in this ideal context is
  \begin{equation}
    \label{eq:cosmicvar}
    \varcosm\ji = 2(4\pi)^{-2}\sum (b_\ell\ji)^4(2\ell + 1)C_\ell^2
  \end{equation}
\end{rem}
Consider now the observation model~(\ref{eq:model}).  Up to the
approximations of Section~\ref{sec:appr-deal-with}, one finds that
\begin{equation}
  \label{eq:hatCj_gal}
  \widehat C\ji 
  :=
  \frac1{\left(B\ji\right)^2}
  \sum_{k \in {\cal K}_{t_j}\ji} {w}_k\ji \left\{
    \left(\gamma_k\ji\right)^2 
    - 
    \left(n_k\ji\right)^2
  \right\}.
\end{equation}
is an unbiased estimate of $C\ji$ as soon as
\begin{equation}
  \label{eq:constraint}
  \sum_{k \in  {\cal K}_{t_j}\ji} w_k\ji = 1.
\end{equation}
The weights can further be chosen to minimize the mean-square error
$\esp\left(\widehat C\ji - C\ji\right)^2$, under the
constraint~(\ref{eq:constraint}). 
It amounts to setting the weights according to the local
signal-to-noise ratio, which is non constant for non stationary noise.
This is a distinctive advantage of our method that it allows for such
a weighting in a straightforward and natural manner.
In the case of uncorrelated coefficients, this optimization problem is
easy to solve (using Lagrange multipliers) and is equivalent to
maximizing the likelihood under the approximation of independent
coefficients (see Appendix~\ref{sec:vari-estim-fusion} for details).
It leads to the solution
\begin{multline}
\label{eq:optimweights}
  w_k\ji(\overline{C}) 
  \\:=
  \left(\overline{C} + \left( n_k\ji\right)^2 \right)^{-2} 
  \Bigl[
  \sum_{k'\in\mathcal{K}\ji_{t_j}}\bigl(\overline{C} +  \left(n_{k'}\ji\right)^2\bigr)^{-2}  
  \Bigr]^{-1}
\end{multline}
with $\overline{C} = C\ji$. This is the unknown quantity to be estimated but it can
be replaced by some preliminary estimate (for example the spectral estimate
of~\cite{Hinshaw+2006}).  One can also iterate the estimation procedure from
any starting point.  The robustness of this method with respect to the prior
spectrum is demonstrated at Section~\ref{sec:choice-parameters} (see
Figure~\ref{fig:compareinit}).

Those weights are derived under the simplifying assumption of independence of
needlet coefficients. They can be used in practice because needlet coefficients
are only weakly dependent. Precisely,  for two fixed points on a increasingly fine grid
$\xi_k, \xi_{k'}$ and well-chosen window functions, the needlets coefficients
$(\gamma_k\ji, \gamma_{k'}\ji)$ are asymptotically independent as $j \to
\infty$ (see \cite{baldi:etal:2008a}). Note that this property is shared by
well-known Mexican Hat wavelets, as proved in \cite{mayeli:2008}.

\subsection{Smooth spectral estimates from multiple experiments}
\label{sec:spectr-estim-multiple}

Consider now the observation model described by Eq.~(\ref{eq:model2}) with
noise independent between experiments.  Using the approximations of
Section~\ref{sec:appr-deal-with}, Eq.~(\ref{eq:model2}) translates to
\begin{equation}
  \label{eq:model2needlet}
  \frac{\gamma_{k,e}\ji(X)}{B_e\ji} 
  =
  \gamma_{k}\ji(T) +
  \frac{n_k\ji(\sigma_e)}{B_e\ji}\; Z_{k,e} 
\end{equation}
in the needlet domain, for indexes $k\in \mathcal{K}\ji_{e,t_j}$, where
$Z_{k,e}$ are standard Gaussian random variables which are correlated within
the same experiment $e$ but independent between experiments.
As explained in the single experiment case of
Section~\ref{sec:spectr-estim-single}, the coefficients are only
slightly correlated.  This justifies the use, in the angular power
spectrum estimator, of the weights derived by the maximization of the
likelihood with independent variables.
The correlation between coefficients does not introduce any bias here
but only causes loss of efficiency.  As in the single-experiment case,
only the coefficients sufficiently far away from the mask of at least
one experiment are kept. Defining
$$
\mathcal{K}\ji = \cup_e \mathcal{K}\ji_{e,t_j}.
$$
the aggregated estimator is implicitly defined (see
Appendix~\ref{sec:vari-estim-fusion}) by
\begin{equation}
  \label{eq:implicitMLfus}
  \widehat C^{ML,(j)}
  =
  \sum_{k\in\mathcal{K}\ji} \tilde w_k\ji\left(\widehat C^{ML,(j)}\right)
  \left\{
    \left(\tilde\gamma\ji_k
    \right)^2
    -
    \left(\tilde n_k\ji\right)^2
  \right\}
\end{equation}
with, for any $k$ in $\mathcal{K}\ji$
\begin{gather}
  \tilde\gamma\ji_k =   \sum_e
  \omega_{k,e}\ji
  \frac{\gamma_{k,e}\ji}{B_e\ji}
  \label{eq:defaggregatedcoef}  
  \\
  \tilde n_k\ji
  :=
  \left[
    \sum_e
    \left(
      \frac{B_e\ji}{n_k\ji(\sigma_e)}
    \right)^2
    \ind_{k\in\mathcal{K}_{e,t_j}\ji}
  \right]^{-1/2} \label{eq:weightsfusioninter}
\\
  \omega_{k,e}\ji 
  := \left(
    \frac{B_e\ji}{n_k\ji(\sigma_e)}
  \right)^2
  \ind_{k\in\mathcal{K}_{e,t_j}\ji}
  \left(\tilde n_k\ji\right)^2\; \label{eq:defomegak}
\end{gather}
and similarly to~(\ref{eq:optimweights}), 
\begin{multline}
\tilde w_k\ji(C)
 \\ :=
  \left(C + \left(\tilde n_k\ji\right)^2 \right)^{-2} 
  \left[ \sum_{k'\in\mathcal{K}\ji}\left(C +  \left(\tilde n_{k'}\ji\right)^2\right)^{-2}  \right]^{-1}
  \label{eq:weightsfusionintra}
\end{multline}
Note that $\sum_e \omega_{k,e}\ji = 1$ and $\sum_k \tilde w_k\ji = 1$.
An explicit estimator is obtained by plugging some previous, possibly
rough, estimate $\overline{C}\ji$ of $C\ji$ in place of $C$ of
Eq.~(\ref{eq:weightsfusionintra}).  
Eventually, the aggregated angular power
spectrum estimator is taken as
\begin{equation}
\label{eq:estimatorFusion}
\widehat C\ji
=
\sum_{k\in \mathcal{K}\ji } \tilde w_k\ji\left(\overline{C}\ji\right)
\left\{
  \left(
    \tilde\gamma\ji_k\right)^2
  -
  (\tilde n_k\ji)^2
\right\}\; .
\end{equation}
This expression can be interpreted in the following way. For any pixel
$k$ in ${\mathcal K}\ji$, that is for any pixel where the needlet
coefficient is reasonably uncontaminated by the mask for at least one
experiment, compute an aggregated needlet coefficient
$\tilde\gamma\ji_k$ by the convex combination
(\ref{eq:defaggregatedcoef}) of the debeamed needlet coefficients from
all available experiments.  
Weights of the combination are computed according to the relative
local signal to noise ratio (including the beam attenuation).
Finally, a spectral estimation is performed on the single map of
aggregated coefficients, in the same way as in
Section~\ref{sec:spectr-estim-single}. 
Those coefficients are squared and translated by $\tilde n\ji_k$ to
provide an unbiased estimate of $C\ji$. Then all the available squared
and debiased coefficients are linearly combined according to their
relative reliability $\tilde w_k\ji(C) $ which is proportional to
$(C\ji + (\tilde n_{k}\ji)^2)^{-1}$.  Figures~\ref{fig:methodFusion1}
and~\ref{fig:methodFusion2} display the values of those weights (maps)
$\tilde w\ji_k$ and $\omega_{k,e}\ji$ for a particular mixing of
experiments.  See Section~\ref{sec:fusi-mult-exper} for details.

\subsection{Parameters of the method}

In this section, we discuss various issues raised by the choice of the
parameters of the NSE method.  Those parameters are: the shape of the
spectral window function $b\ji_\ell$ in each band (or equivalently the
shape of the needlet itself in the spatial domain), the bands
themselves (\emph{i.e.}  the spectral support of each needlet) and the
values of the thresholds $t_j$ that define the regions of the sky
where needlets coefficients are trusted in each band; see
Eq.~(\ref{eq:Wt}).
See Section~\ref{sec:choice-parameters} for a numerical investigation.

\subsubsection{Width and shape of the window functions}
\label{sec:shap-wind-funct}

For spectral estimation, it is advisable to consider spectral window functions
with relatively narrow spectral support, in order to reduce bias in the
spectral estimation. The span of the summation in (\ref{eq:defneedlet}) can be
fixed to some interval $[\lmin\ji, \lmax\ji]$. For our illustrations, the
interval bands have been chosen to cover the range of available multipoles with
more bands around the expected positions of the peaks of the CMB. The bands are
described in Table~\ref{tab:bands}.

\begin{table*}
  \centering
  \setlength{\tabcolsep}{3pt}.
  {\small
  \begin{tabular}{|c|c|c|c|c|c|c|c|c|c|c|c|c|c|c|c|c|c|c|c|c|c|c|c|c|c|c|c|c|c|c|c|c|c|c|c|c|c|c|c|c|c|c|c|c|c|c|}
    \hline
    Band $(j)$ & 1   & 2   & 3   & 4   & 5   &$\cdots$& 20  & 21  & 22  & 23  & 24  & 25  & 26  & 27  & 28  & 29  &$\cdots$& 35  & 36  & 37  & 38  & 39\\
    \hline \hline
    $\lmin\ji$ & 2   & 11  & 21  & 31  & 41  &\multirow{2}{*}{$\cdots$}        & 401 & 451 & 501 & 551 & 601 & 651 & 701 & 751 & 801 & 876 &  \multirow{2}{*}{$\cdots$}      & 1326& 1426& 1501& 1626& 1751\\ 
    $\lmax\ji$ & 20  & 30  & 40  & 50  & 60  &        & 500 & 550 & 600 & 650 & 700 & 750 & 800 & 875 & 950 & 1025&        & 1475& 1625& 1750& 1875& 2000\\
    \hline \hline
    $\nside{}\ji$& 16  & 16  & 32  & 32  & 32  & $\cdots$  & 256 & 512 & 512 & 512 & 512 & 512 & 512 & 512 & 512 & 1024& $\cdots$ & 1024& 1024& 1024& 1024& 1024\\
    \hline \hline
  $\theta_0\ji$ 
    &   69 &  50 &  41  & 36
    & 32& $\cdots$& 10.7&10.2& 9.7& 9.3& 8.9& 8.6& 8.3& 8.0& 7.7& 7.4
    &$\cdots$&   6.1 &5.8 & 5.6 & 5.4 & 5.2\\
\hline
  \end{tabular}
}
\caption{Spectral bands used for the needlet decomposition in this
  analysis. Depending on $\lmax\ji$, the needlet coefficient maps are computed
  using the HEALPix package, at different values of \nside, given in the fourth
  line. The number  $K\ji$ of needlet coefficients in band $j$ is then $12(\nside\ji)^2$. It is a kind of  \emph{decimated} implementation of the
  needlet transform. The last line gives the opening $\theta\ji$ (in degrees)
  chosen in Eq.~\ref{eq:optimprob1} to
  define the PSWF from the \emph{Prolate 2} family (see
  Section~\ref{sec:choice-parameters} for details).}
  \label{tab:bands}
\end{table*}

\begin{figure*}
  \centering
  \includegraphics[width=0.49\linewidth]{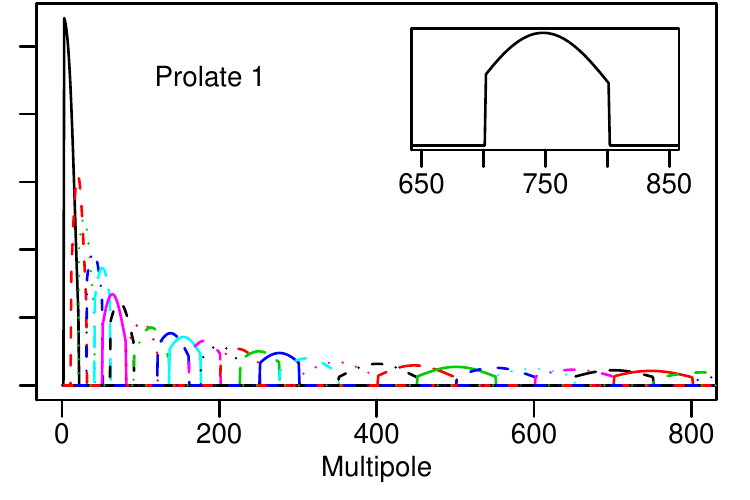}
  \includegraphics[width=0.49\linewidth]{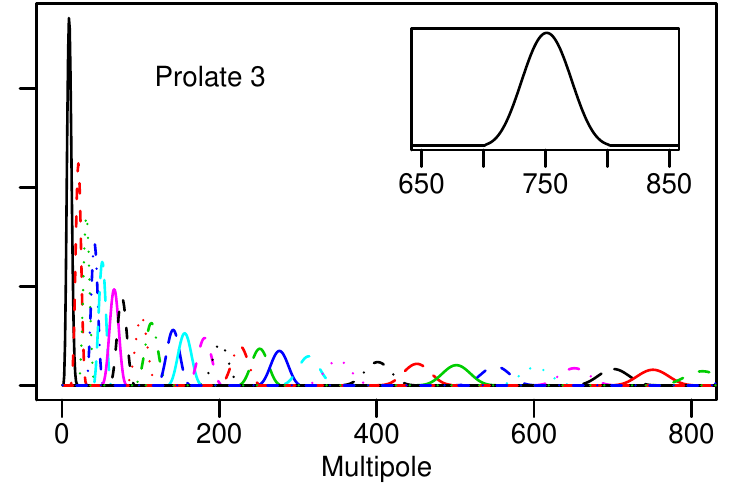}\\
  \includegraphics[width=0.49\linewidth]{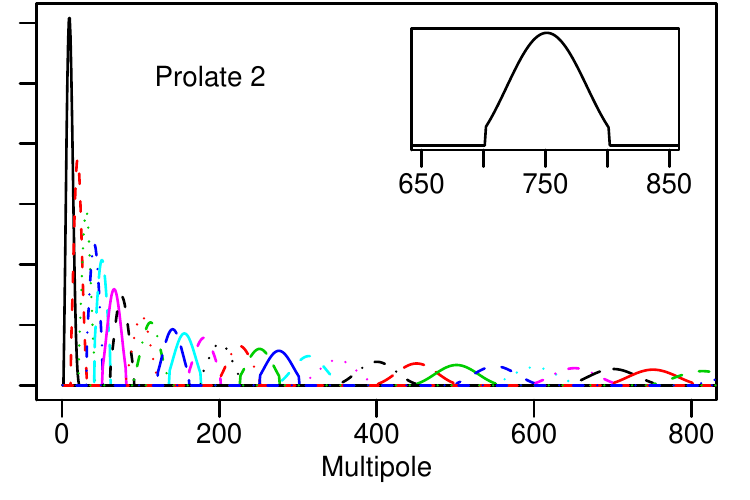}
  \includegraphics[width=0.49\linewidth]{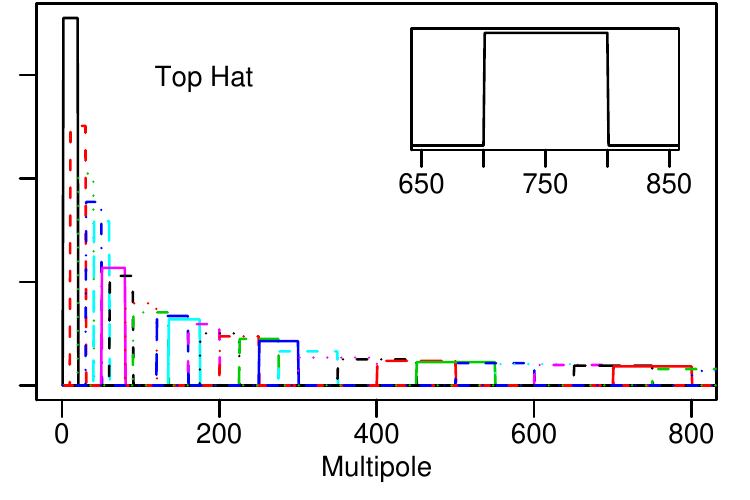}\\
  \caption{Four families of window functions that are used for the NSE and
    compared numerically in Section~\ref{sec:choice-parameters}. There are
    three families of prolate spheroidal wave functions and one family of
    top-hat functions. All the families are defined on the same bands. Inset
    graphs show the window function in the 26th band. Each window function is
    normalized by the relation $(4\pi)^{-1} \sum (b_\ell\ji)^2(2\ell+1) =
  1$. Then, if the angular power spectrum is flat, $C_\ell  \equiv C_0$, then
  $C\ji \equiv C_0$ for all bands, according to~(\ref{eq:defcj}).} 
  \label{fig:windowfunctions}
\end{figure*}

\begin{figure*}
  \centering
  \includegraphics[width=0.8\linewidth]{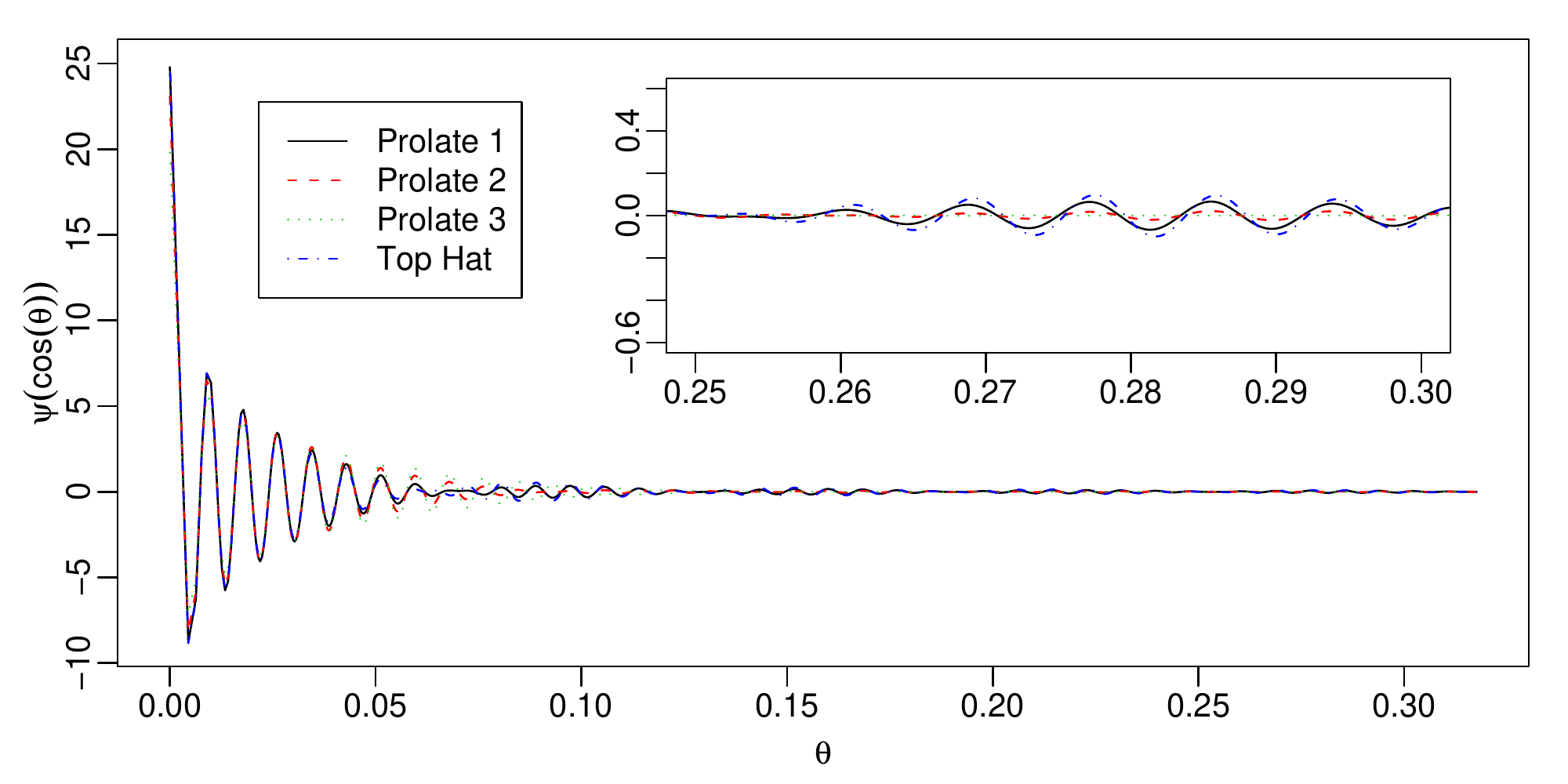}
  \caption{Angular profile of the four needlets associated to the window functions at
    the 26th band ($701 \leq \ell \leq 800$) from the four families of
    Figure~\ref{fig:windowfunctions}. We have plotted the axisymmetric profile
    $\sum_\ell b_\ell L_\ell (\cos \theta)$ as a function of $\theta$. Needlets
    with ``smoother'' associated window profile (such as Prolate 3) need more
    room to get well localized, but are less bouncing than needlets with abrupt
    window function (such as top hat or Prolate 1)}
  \label{fig:needlets}
\end{figure*}

It is well known however that perfect spectral and spatial localization cannot
be achieved simultaneously (call it the uncertainty principle). In order to
reduce the effect of the mask, we have to check that the analysis kernels are
well localized. This leads to the optimization of some localization
criteria. If we retain the best $\Lset^2$ concentration in a polar cap
$\Omega_{\theta\ji}=\{\xi:\theta\leq\theta\ji\}$, namely
\begin{multline}
  \label{eq:optimprob1}
  (b\ji_{\ell})_{\ell = \lmin,\cdots,\lmax} \\= \arg\max_{\textsf b}
  \frac{\int_{\Omega_{\theta\ji}} \left| \sum_{\ell=\lmin\ji}^{\lmax\ji}
      {\textsf b}\ji_\ell L_\ell (\xi) \right|^2 \dd\xi}{\int_{\Sset} \left| \sum_{\ell=\lmin\ji}^{\lmax\ji}
      \textsf{b}\ji_\ell L_\ell (\xi) \right|^2 \dd\xi}
\end{multline}
we obtain the analogous of prolate spheroidal wave function (PSWF)
thoroughly studied in \eg Slepian \& Pollak (1960)\nocite{Slepian+60}
for the PSWF in $\Rset$ and
\cite{simons:dahlen:wieczorek:2006,guilloux:fay:cardoso:2008} for PSWF
on the sphere.   
In our simulations, we use PSWF needlets since they are well localized
and easy to compute.  Other criteria and needlets can be investigated
and optimized, at least numerically; see
\cite{guilloux:fay:cardoso:2008} for details.  
The choice of the optimal window function in a given band is a non
trivial problem which involves the spectrum itself, the
characteristics of the noise and the geometry of the mask.  Even if we
restrict to PSWF as we do here, it is not clear how to choose the
optimal opening $\theta\ji$ for each band $j$.  We can use several
rules of thumb based on approximate scaling relation between roughly
$B$-adic bands and openings $\theta\ji$ that preserve some Heisenberg
product or Shannon number.  
Figure~\ref{fig:windowfunctions} represents three families of PSWF
needlets that are numerically compared below.  Their spatial
concentration is illustrated by Figure~\ref{fig:needlets}.

\subsubsection{The choice of the needlets coefficients (mask)}
\label{sec:choice-needl-coeff}

Practically we want to keep as much information (\ie as many needlet
coefficients) as possible, and to minimize the effect of the mask.  Using all
the needlet coefficients regardless of the mask would lead to a biased estimate
of the spectrum. It is still true if we keep all the coefficients outside but
still close to the mask, keeping in mind that the needlets are not perfectly
localized. On the other hand, getting rid of unreliable coefficients reduces
the bias, but increases the variance. This classical trade-off is taken by
choosing the threshold level $t\ji$ in the Definition~(\ref{eq:Wt}) of the
excluding zones. For multiple experiments, a different selection rule can be
applied to each experiment, according to the geometry of the mask and the
characteristics of the beam and the noise.

\section{Monte Carlo studies}
\label{sec:monte-carlo-studies}

Recall that NSE spectral estimators are designed based on three
approximations:
\begin{itemize}
\item one can neglect the impact of the mask on the needlet
  coefficients which are centered far enough from its edges;
\item one can neglect the variations of the beam and the CMB power
  spectrum over each band.
\item the weights, which are optimal under the simplifying assumption of
  independent needlet coefficients, still provide good estimates for the truly
  weakly correlated needlet coefficients.
\end{itemize}
We carry out Monte Carlo studies to investigate, first the actual performance
of the method on realistic data, and second the sensitivity of the method with
respect to its parameters. Stochastic convergence results under appropriate
conditions is established in a companion paper \cite{fay:guilloux:2008}.

\subsection{Single map with a mask and inhomogeneous noise}
\label{sec:mask-inhom-noise}

In this section, we first consider model~(\ref{eq:model}).  According
to Eqs~(\ref{eq:varneedcoeffdebeam}) and~(\ref{eq:hatCj_gal}), any
beam can be taken into account easily in the procedure.  Without loss
of generality, we suppose here that there is no beam (or $B$ is the
Dirac function).  The case of different beams is addressed in
Section~\ref{sec:fusi-mult-exper}, see Table~\ref{tab:fusionparams}.

\begin{figure*}
  \def\widthfig{0.23\linewidth}
  \centering
  \subfigure[Mask]{
    \includegraphics[width=\widthfig]{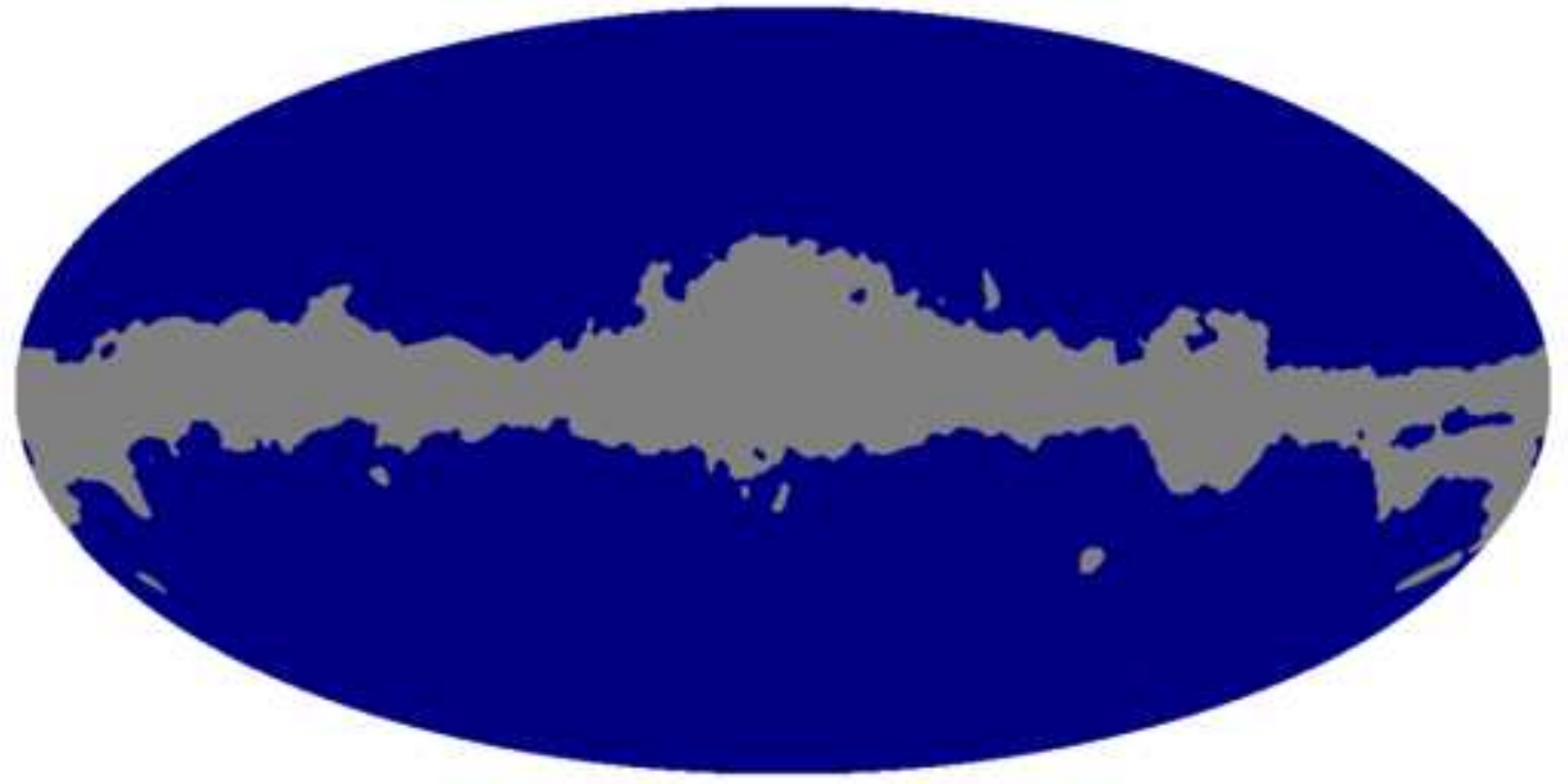}
  }
  \subfigure[Noise standard deviation]{
    \includegraphics[width=\widthfig]{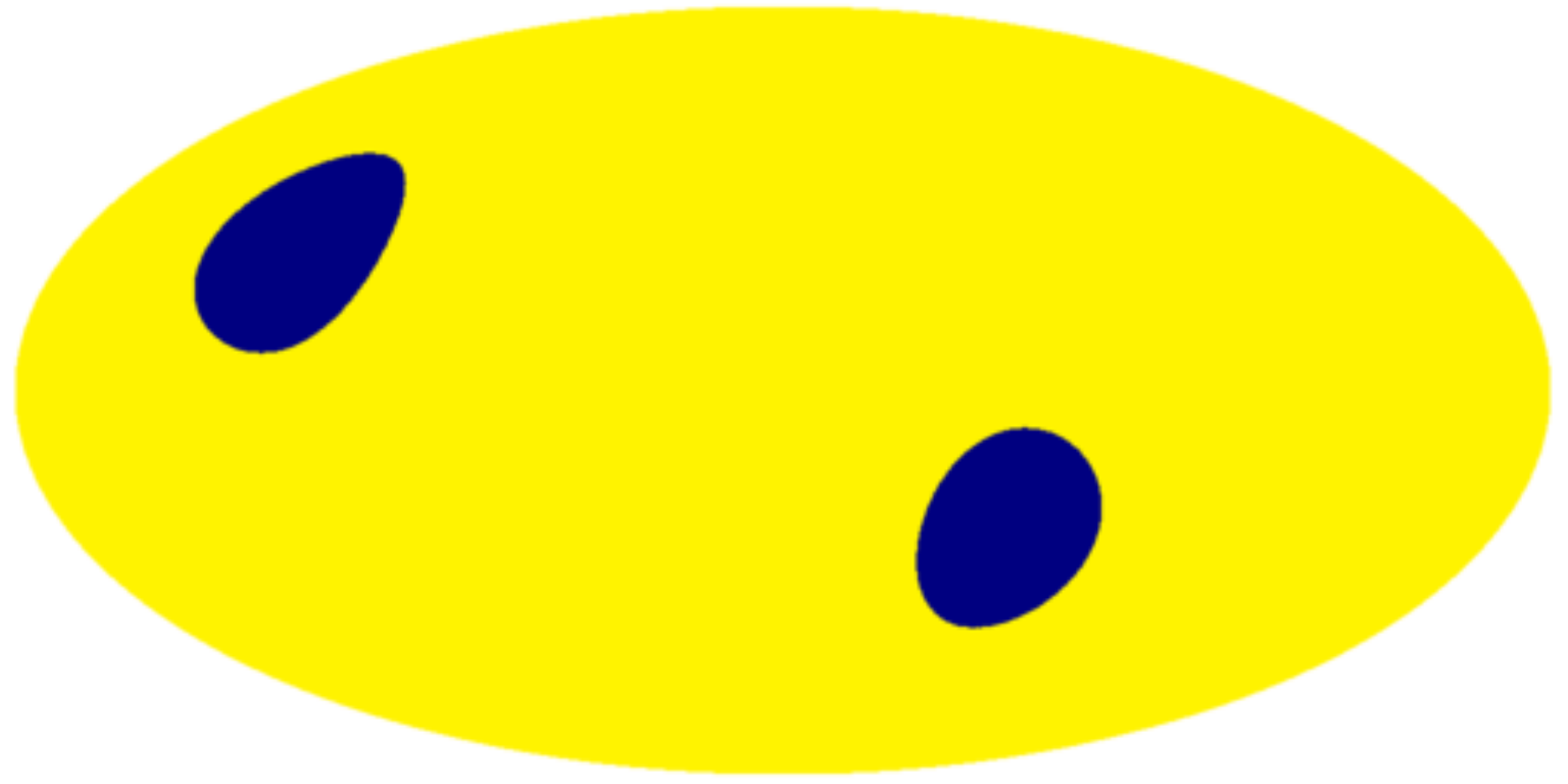}
  }
  \subfigure[One simulated input map]{
    \includegraphics[width=\widthfig]{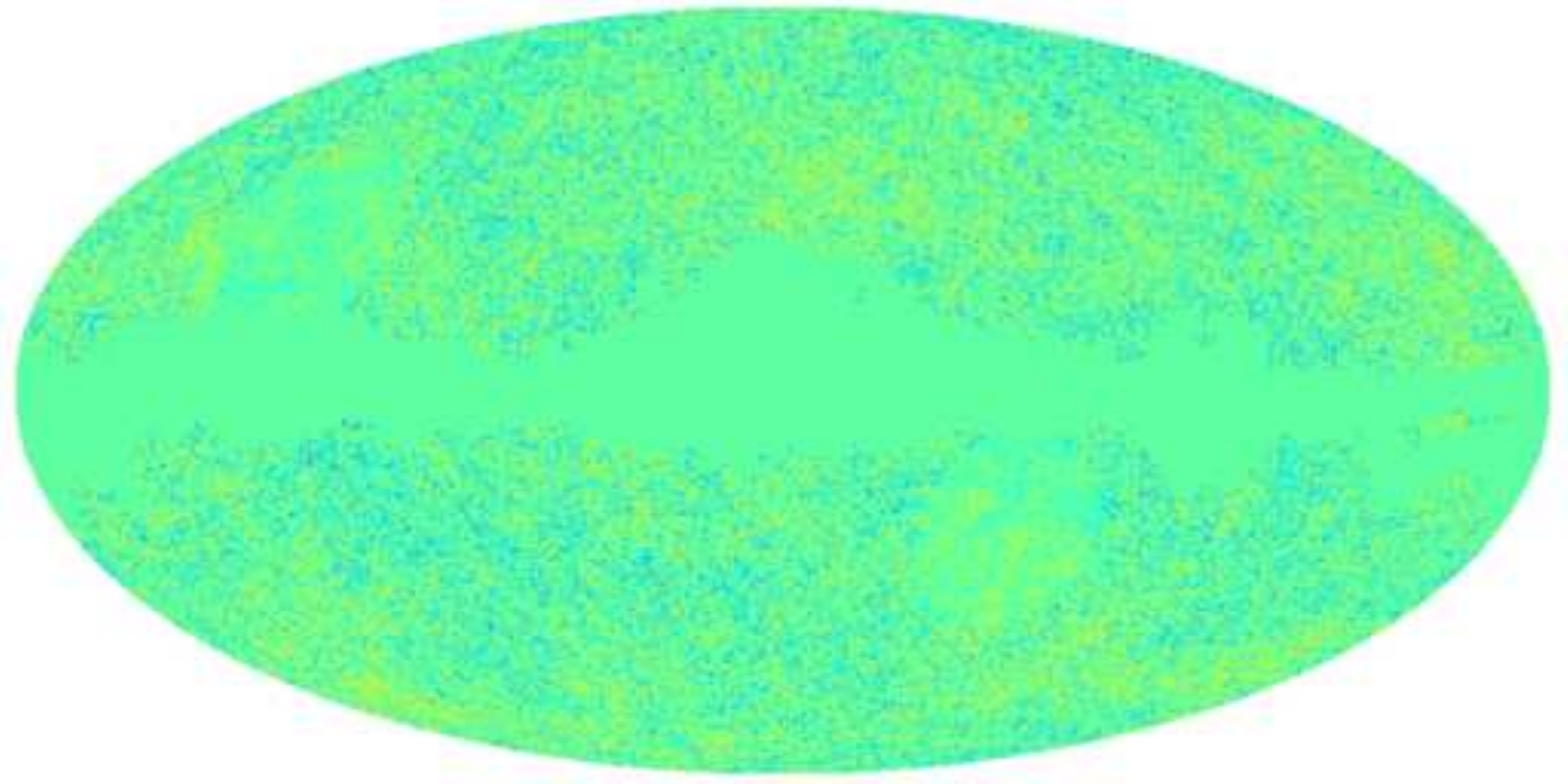}
  }
  \subfigure[Spectra]{
    \includegraphics[width=\widthfig]{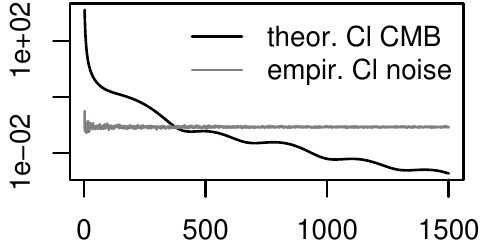}
  }
  \caption{Simplified model of partially covered sky and inhomogeneous additive
    noise. This model is used to compare numerically the NSE estimator with PCL
    estimators and to assess the robustness or the sensitivity of the method
    with respect to its parameters. The mask is kp0. In CMB $\mu K$ units, the standard
    deviation of the uncorrelated pixel noise is 75 in the two small circular patches and 300 elsewhere.}
  \label{fig:toy1}
\end{figure*}

The key elements for this numerical experiment are illustrated by
Fig.~\ref{fig:toy1}.  We simulate CMB from the spectrum $C_\ell$ given
by the $\Lambda$CDM model that best fits the W-MAP data.  We use a Kp0
cut \citep{bennett:2003basic} for the mask and we take a simple non
homogeneous noise standard deviation map (the SNR per pixel is 1.5 in
two small circular patches and 0.4 elsewhere).

Using a mean-square error criterion, we first study the dependence of
NSE performance on its free parameters.  Then we compare NSE with
methods based on Spherical Harmonic coefficients, known as
pseudo-$C_\ell$ estimation and followed in Hinshaw {\it et al.}
(2006)\nocite{Hinshaw+2006}). 
For the reader's convenience, the PCL procedure is summarized in
Appendix~\ref{sec:pseudo-c_ell-estim}.

\subsubsection{Mean-square error}
\label{sec:mean-square-error}

We shall measure the quality of any estimator $\widehat C\ji$of $C\ji$ by its
mean-square error \[
\text{MSE}(\widehat C\ji) = \esp \left(\widehat C\ji- C\ji\right)^2 \ .
\]
This expectation is estimated using \nmc{} Monte Carlo replications. Roughly
speaking, the MSE decomposes as an average estimation error and a
\emph{sampling variance}.  The estimation error term is intrinsic to the
method. Ideally, it should be used to compare the relative efficiency of
concurrent approaches.  The sampling variance term is the so-called
\emph{cosmic variance}. It is given by the characteristic of the spectrum and
coming from the fact that we only have one CMB sky, and thus $2\ell+1$ $a\lm$'s
to estimate one $C_\ell$. It is increased by the negative influence of the
noise and the mask. This gives an error term intrinsic to the whole experiment.
When the sky is partially observed (let \fsky{} denote the fraction of
available sky) and for high $\ell$'s (or $j$'s), the cosmic variance must be
divided by a factor \fsky{} leading to the following approximate Cram\'er-Rao
lower bound at high frequencies
\begin{equation}
  \label{eq:cosmicvarfsky}
  \varsample\ji =  \fsky^{-1} \varcosm\ji \ .
\end{equation}
 Including an homogeneous additive uncorrelated pixel noise with
variance $\sigma^2$, the sample variance writes
\begin{equation*}
  2 \fsky^{-1}\sum (b_\ell\ji)^4(2\ell + 1) \left(C_\ell + \frac{4\pi}\npix \sigma^2\right)^2
\end{equation*}
In a non-homogeneous context, no close expression for the sampling
variance is available: Eq.~(\ref{eq:cosmicvarfsky}) will serve as one
reference.
When comparing different window functions in the same band, it must be kept in
mind that different estimators do not estimate the same $C\ji$ so that the
sampling variances are not the same.  In this case, we use the following
normalized MSE
\begin{equation}
  \label{eq:normalizedMSE}
  \frac{\text{MSE}(\widehat C\ji)}{\fsky^{-1}\varcosm\ji} 
\end{equation}

\subsubsection{Robustness with respect to parameter choice}
\label{sec:choice-parameters}

This section looks into the robustness of NSE with respect to its free
parameters.

\begin{figure}
  \centering
  \includegraphics[width=\linewidth]{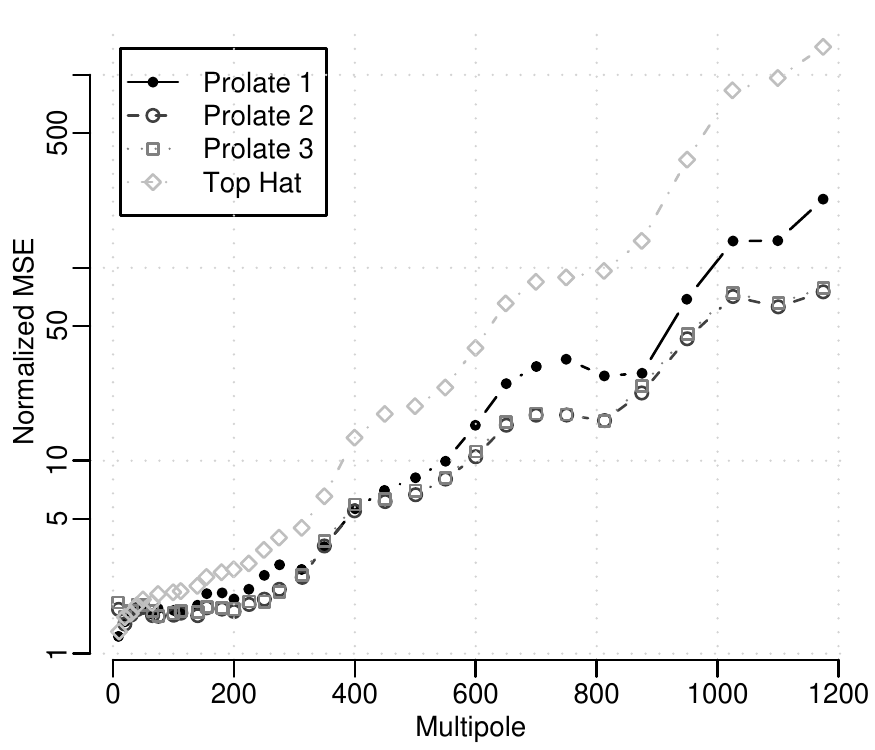}
  \caption{Comparison of the normalized MSE~(\ref{eq:normalizedMSE}) of the
    needlet spectral estimators for the four  families of spectral window
    functions displayed in Figure~\ref{fig:windowfunctions}. The smoothness of the window function make the MSE smaller at high
    multipoles. At low multipoles, taking a too smooth function makes the
    needlet less localized and there is a loss of variance due to the smaller
    number ${\cal K}\ji$ of needlet coefficients that are combined.}
  \label{fig:comparefam}
\end{figure}

First and as expected, the spectral estimation is very sensitive to
the choice of the window functions.  Even if we restrict to the PSWF,
one has the freedom to choose a concentration radius $\theta\ji$ for
each band.  We compare the mean-square error of the estimation for
various choices of $\theta\ji$ that lead to three of the window
function families displayed in Figure~\ref{fig:windowfunctions}.  The
second prolate family is obtained using the ``rule of thumb'' relation
$\theta\ji = 2((\lmin\ji + \lmax\ji)/2)^{-1/2}$.  The values of those
opening angles are in Table~\ref{tab:bands}.  The first and third
sequences of opening angles are the same with a multiplicative factor
of 0.5 and 2, respectively.  For the sake of comparison we also
consider the top-hat window functions.  Figure~\ref{fig:comparefam}
shows the normalized MSE for those four ``families'' of needlets as a
function of the band index.
Notice the poor behavior of a non-optimized window function and the far better
performance of the second prolate family in comparison with top-hat and Prolate
1 windows. Thus, in the following, we use this particular needlet family to
study the sensitivity of the method with respect to the other parameters, and
to compare NSE and PCL estimators.

\begin{figure}
  \centering
  \includegraphics[width=\linewidth]{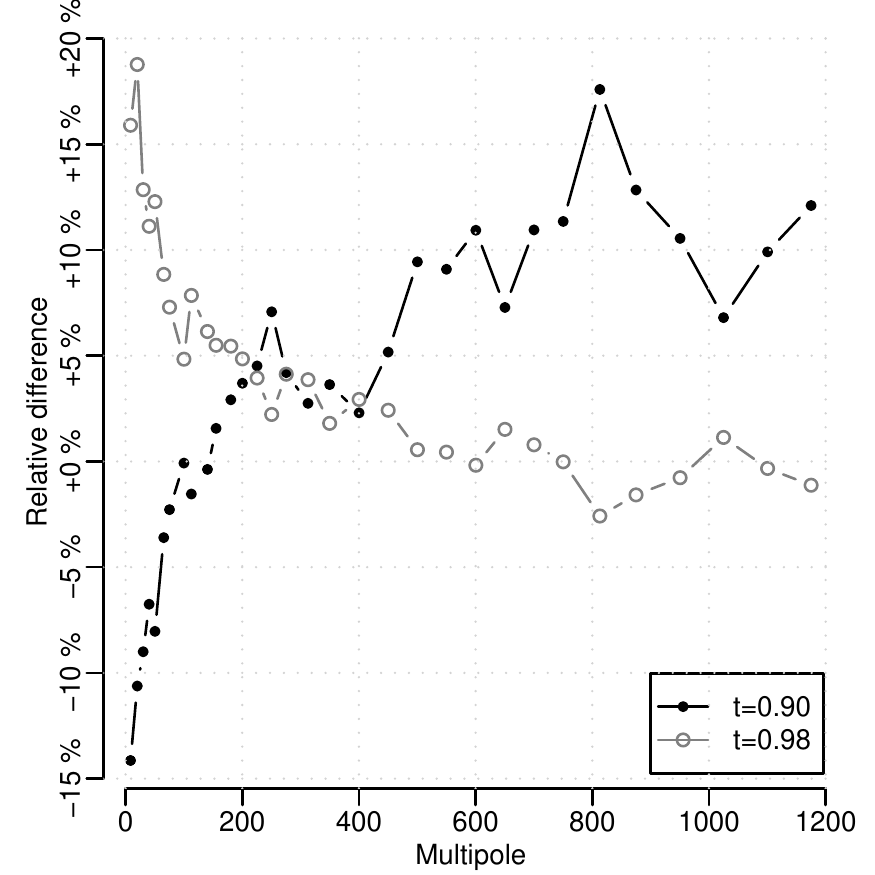}
  \caption{Relative difference between the normalized
    MSE~(\ref{eq:normalizedMSE}) of the needlet spectral estimation using
    thresholds 0.9 and 0.98, and the same with threshold 0.95. It
    highlights the fact that the estimation is not very sensitive to the value
    of this parameter, except at low $\ell$'s, where we do not advocate the use
  of the NSE. The window function family is ``Prolate 2'' from Figure~\ref{fig:windowfunctions}.}
  \label{fig:diffthresh}
\end{figure}

Next, for the second family of PSWF, we compare the influence of the
threshold value $t\ji \equiv t$ for $t=0.9, 0.95$ or $0.98$.
Figure~\ref{fig:diffthresh} shows that this choice within reasonable
values is not decisive in the results of the estimation procedure. 
For very low frequencies ($\ell \leq 100$), the conservative choice
$t=0.98$ increases the variance since many needlets are contaminated
by the mask and discarded.  Qualitatively, in such variance dominated
regimes, taking more coefficients (\eg $t = 0.9$) is
adequate. However, we do not advocate the use of the NSE at low
$\ell$'s where exact maximum likelihood estimation is doable.  At
higher $\ell$'s, there is roughly no difference between the $t=0.95$
and $t=0.98$ thresholds.

\begin{figure}
  \centering
  \includegraphics[width=\linewidth]{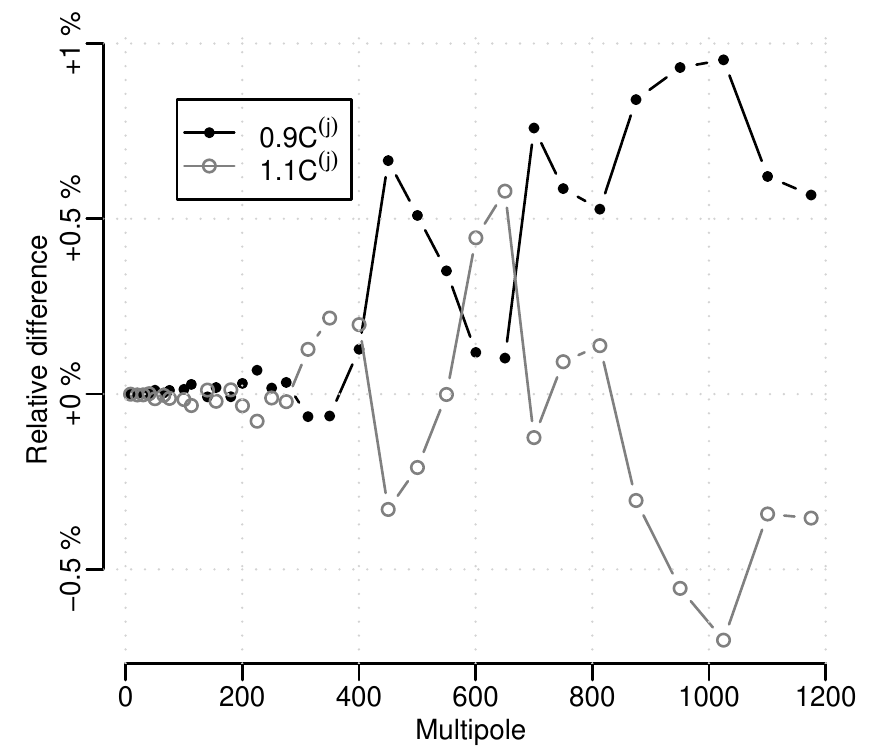}
  \caption{Robustness of the NSE with respect to the initial value $\bar C\ji$
    given to the weights formula~(\ref{eq:optimweights}). We have performed the
    whole estimation with $\bar C\ji = 0.9 C\ji$ and $\bar C\ji = 1.1
    C\ji$. This plot shows the relative difference between the normalized MSE
    under those initial values and the normalized MSE under the optimal initial
    value $\bar C\ji = C\ji$.}
  \label{fig:compareinit}
\end{figure}

Finally, we check the robustness of the method against an imprecise
initial spectrum. 
We take $0.9C\ji$ and $1.1C\ji$ as initial values $\bar C\ji$ and
compare the results with the best possible initial value which is
$C\ji$ itself. The relative difference between the results, displayed
in Figure~\ref{fig:compareinit}, does not exceed 1\%.

\subsubsection{Pseudo-$C_\ell$ versus needlet spectral estimator}
\label{sec:pseudo-cl-vs-NSE}

\begin{figure*}  
  \centering
 \includegraphics[width=0.7\linewidth]{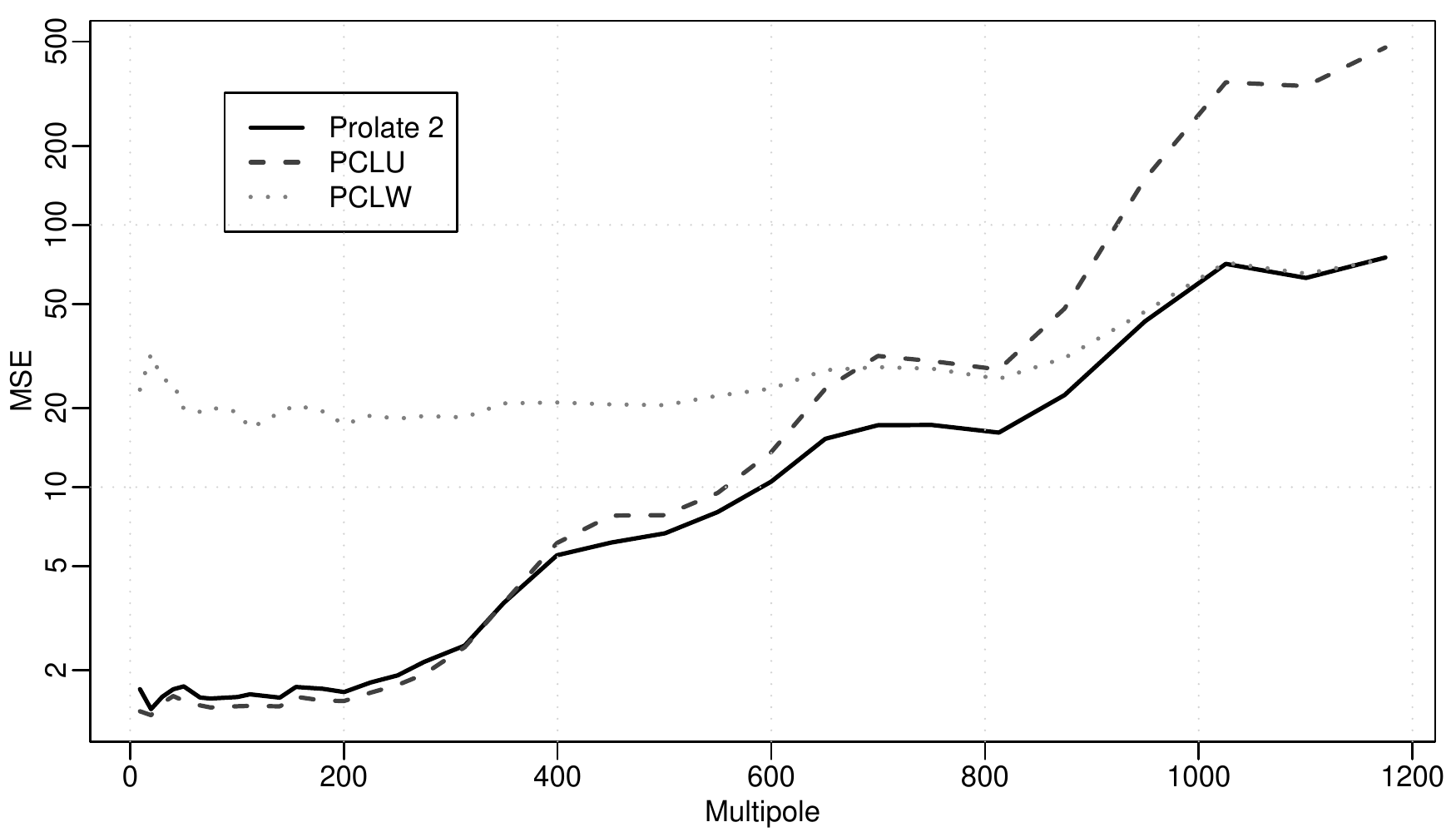}
 \caption{Comparison of the relative MSE (\ref{eq:normalizedMSE}) of the two
   PCL estimators (PCLU for flat weights, PCLW for inverse variance weights)
   with the NSE, for Prolate family 2. For $300\leq \ell \leq 1200$, the NSE is
   uniformly better than the best of the two PCL methods. It should be noted
   that the NSE may be improved again by optimizing the window profiles and the
   thresholds $t\ji$ (\eg by taking a lower threshold for low bands to reduce
   the variance, see Figure~\ref{fig:diffthresh}).}
\label{fig:compNeedPCL}
\end{figure*}

We compare the NSE estimator given by Eq.~(\ref{eq:hatCj_gal}) with
estimation based on the spherical Harmonic coefficients of the
uniformly weighted map and of the $1/\sigma^2$-weighted map.  The
result is displayed in Fig.~\ref{fig:compNeedPCL}.

As expected, at low multipoles, where the SNR is higher, the uniform
pseudo-$C_\ell$ estimator performs better than the weighted pseudo-$C_\ell$
estimator and, conversely, at high multipoles where the SNR is
lower. \cite{efstathiou:2004} proved that the equal-weights pseudo-$C_\ell$
estimator is asymptotically Fisher-efficient when $\ell$ goes to infinity.  The
behaviour of the needlet estimator is excellent: its performance is comparable
to the best of the two previous methods both at low and high multipoles.  Thus,
there is no need to choose arbitrary boundaries between frequencies for
switching for one weighting to the other.  The NSE estimator automatically
implements a smooth transition between the two regimes and it does so
quasi-optimally according to noise and mask characteristics. At low $\ell$'s
one should optimize the window function (the characteristic angle of opening of
the prolates) to broaden the range of optimality of the NSE.

\begin{figure*}
  \centering
  \includegraphics[width=0.35\linewidth]{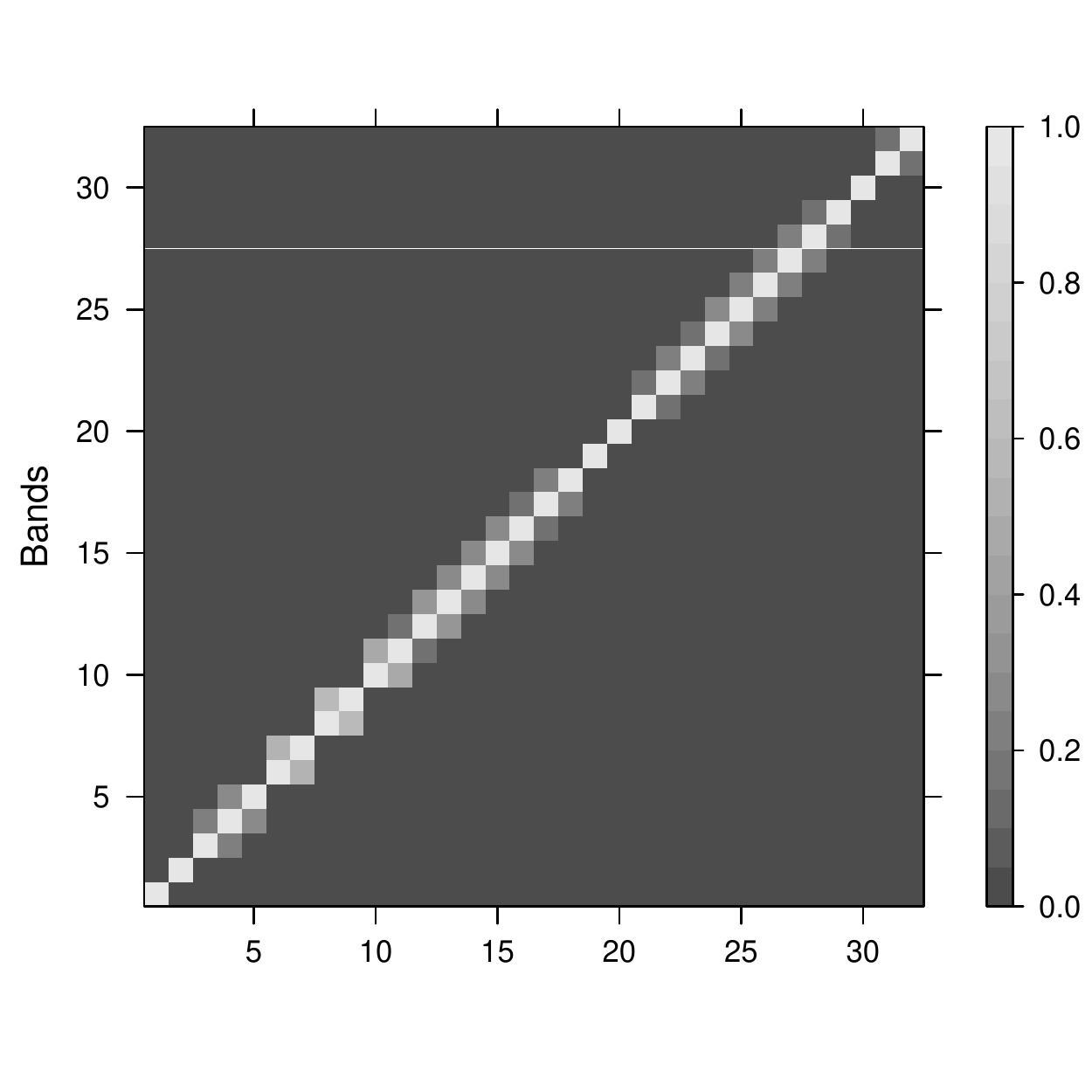}
  \includegraphics[width=0.35\linewidth]{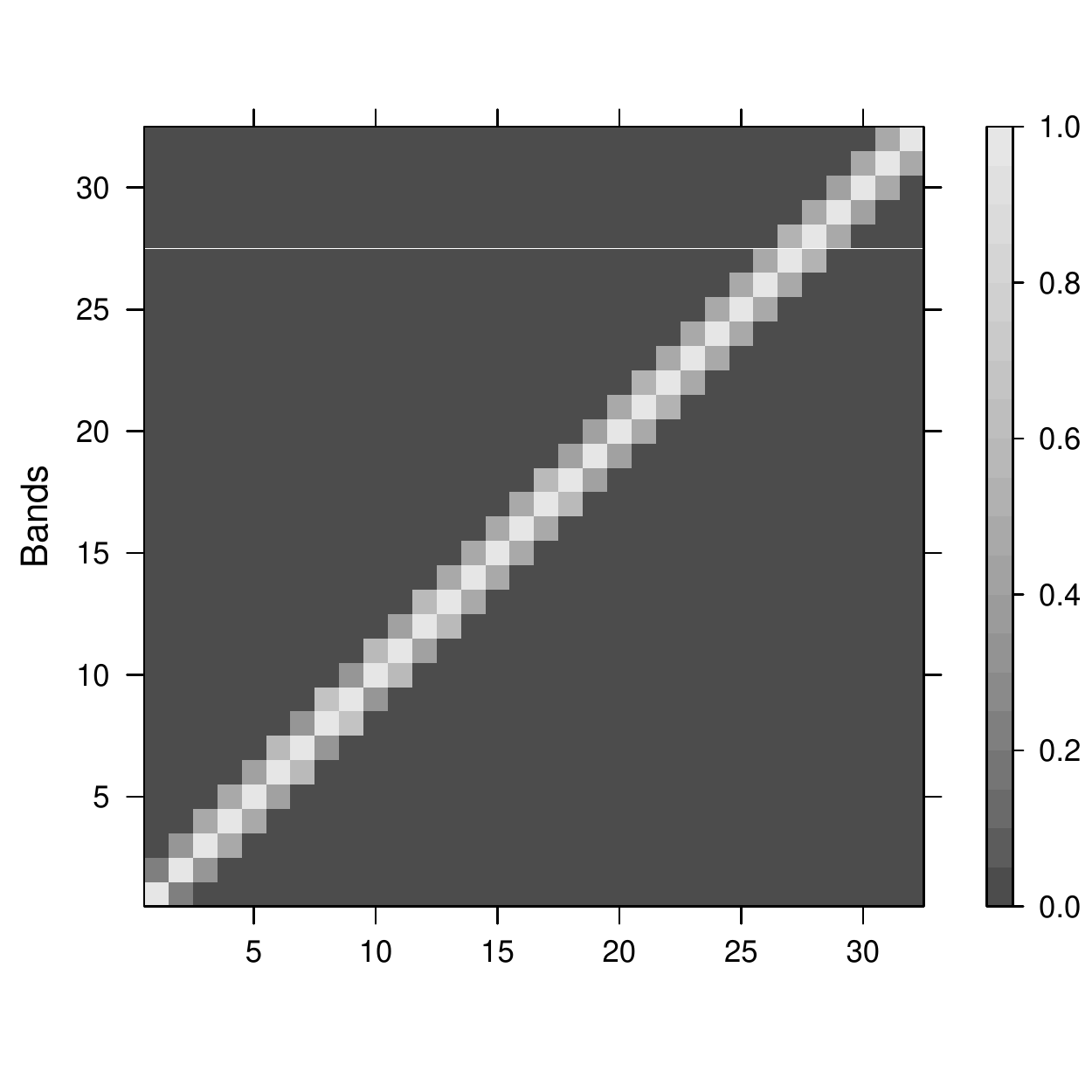} 
\vspace{-1cm}
\caption{Absolute value of the correlation matrix of vector $(\widehat C^{(1)},
  \ldots, \widehat C^{(32)})$, which entries are defined by
  Eq.~(\ref{eq:hatCj_gal}). It has been estimated in the context of
  Fig.~\ref{fig:toy1} using two different families of window functions and
  \nmc{} Monte Carlo replicates. This shows the difference between a family of
  PSWF (left panel) and a family of top-hat windows (right panel).}
  \label{fig:covmat}
\end{figure*}

Providing a $C_\ell$ estimate with error bars is often not sufficient.
Estimate the covariance matrix of the whole vector of spectral estimates is
necessary for full error propagation towards, say, estimates of cosmological
parameters.  Figure~\ref{fig:covmat} shows the values of the correlation matrix
between the spectral estimates.  In the idealistic case of a full sky noiseless
experiment, the theoretical correlation matrix is tridiagonal because window
functions we have chosen only overlap with their left and right nearest
neighbours.  The mask induces a spectral leakage, which is however reduced for
the smoothest window function. This leakage is however compensated for by the
selection of coefficients in $\mathcal{K}\ji_{t\ji}$ (see Eq~(\ref{eq:Wt})).

\subsection{Aggregation of multiple experiments}
\label{sec:fusi-mult-exper}

Historically in CMB anisotropy observations, no single instrument provides the best measurement everywhere on the sky, and for all possible scales.
In the early 90's, the largest scales have been observed first by COBE-DMR, complemented by many ground-based and balloon-borne measurements at higher $\ell$.
Similarly, ten years later, WMAP full sky observations on large and intermediate scales have been complemented by small scale, local observations of the sky as those of 
Boomerang, Maxima, ACBAR or VSA.

The joint exploitation of such observations has been so far very basic. The best power spectrum is obtained by choosing, for each scale, the best measurement available, and discarding the others. 
One could, alternatively, average the measurements in some way, but the handling of errors is complicated in cases where a fraction of the sky is observed in common by more than one experiment.

Clearly, the data is best used if some method is devised that allows combining such complementary observations in an optimal way.
In this section we present the results of a Monte Carlo study to illustrate the
benefits of our method of aggregated spectral estimator.

We simulate observations following the model~(\ref{eq:model2}), with $E=6$
observed maps :
3 Kp0-masked maps with beams and noise-level maps according to W-MAP experiment
in bands Q, V and W respectively ;
3 maps with uniform noise, observed in patches the size of which are equivalent
to BOOMERanG-Shallow, BOOMERanG-Deep and ACBAR observations respectively, and
noise levels representative of the sensitivities of those experiments.
Table~\ref{tab:fusionparams} gives the key features of these experiments.
\begin{table}
  \centering
  \begin{tabular}{|c|c|c|c|c|}
\hline    Experiment & Beam & \texttt{nside} & Noise level & \fsky{}\\
\hline\hline W-MAP Q& 31' &  \multirow{3}{*}{512}  & \multirow{3}{*}{
  \begin{minipage}[c]{2cm}
   \begin{center}{Given by the hit map}
   \end{center}
 \end{minipage}
}
&\multirow{3}{*}{78.57\,\%} \\
\cline{1-2} W-MAP V& 21' & & &\\
\cline{1-2}  W-MAP W& 13' & &&\\
\hline BOOM S&  \multirow{2}{*}{10'} & \multirow{2}{*}{1024} &17.5 $\mu $K & 2.80\,\%\\
\cline{1-1}\cline{4-5} BOOM D && & 5.2 $\mu $K & 0.65\,\%\\
\hline ACBAR & 5' & 2048\footnote{We used \nside=1024 for our Monte Carlo simulations, as
  going to $\lmax \simeq 2000$ is enough to discuss all the features of
  our method.} & 14.5 $\mu $K & 1.62\,\%\\
\hline
  \end{tabular}
  \caption{Main parameters of the experiments to be aggregated. The beams are
    given in minutes of arc, $\nside{}$ refers to HEALPix resolution of the
    simulated maps, noise level is either a map computed from a hitmap and an
    overall noise level, or a uniform noise level per pixel (in $\mu$K
    CMB). Numbers quoted here are indicative of the typical characteristics of
    observations as those of W-MAP, BOOMERanG and ACBAR, and are used for
    illustrative purposes only.}
  \label{tab:fusionparams}
\end{table}
Further details can be found in \cite{bennett:2003basic} and
\cite{hinshaw:2008} for W-MAP, \cite{masi:2006} for BOOMERanG,
\cite{runyan:etal:2003} and \cite{reichardt:etal:2008} for ACBAR.  However, we
do not intend to produce fully-realistic simulations. Basically, no foregrounds
are included in simulations (neither diffuse nor point sources) ; for ACBAR
only the 3 sky fields of year 2002 are used ; and for W-MAP only one detector is
used for each band.

\begin{figure*}  
  \begin{minipage}{0.03\linewidth}
  \includegraphics[width=\linewidth]{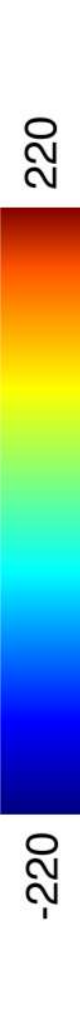}
  \end{minipage}
  \begin{minipage}{0.94\linewidth}
    \def\widthfig{0.23\linewidth}
    \centering
  \subfigure[``True'' CMB]{
    \includegraphics[width=\widthfig]{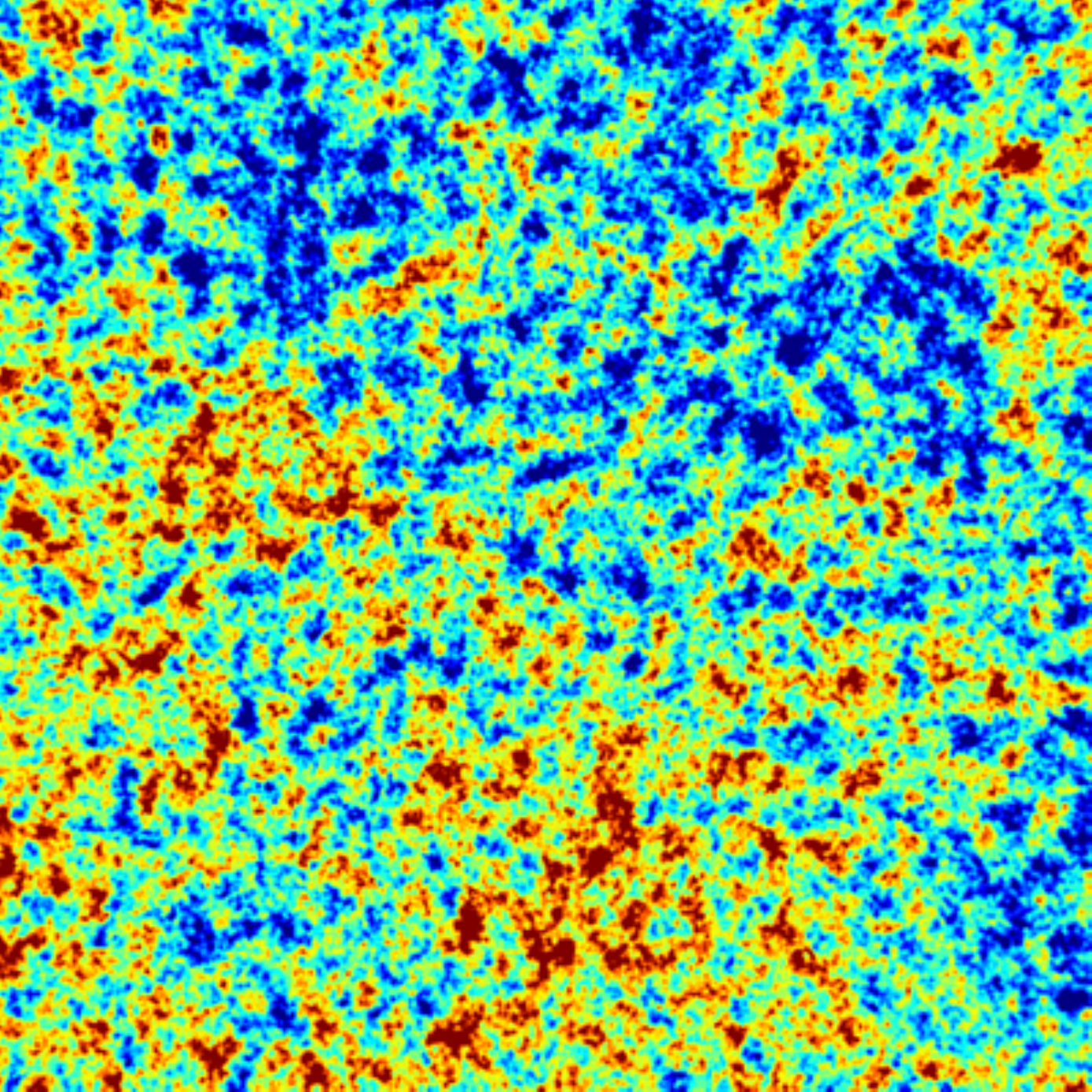}
  }
  \subfigure[WMAP-Q]{
    \includegraphics[width=\widthfig]{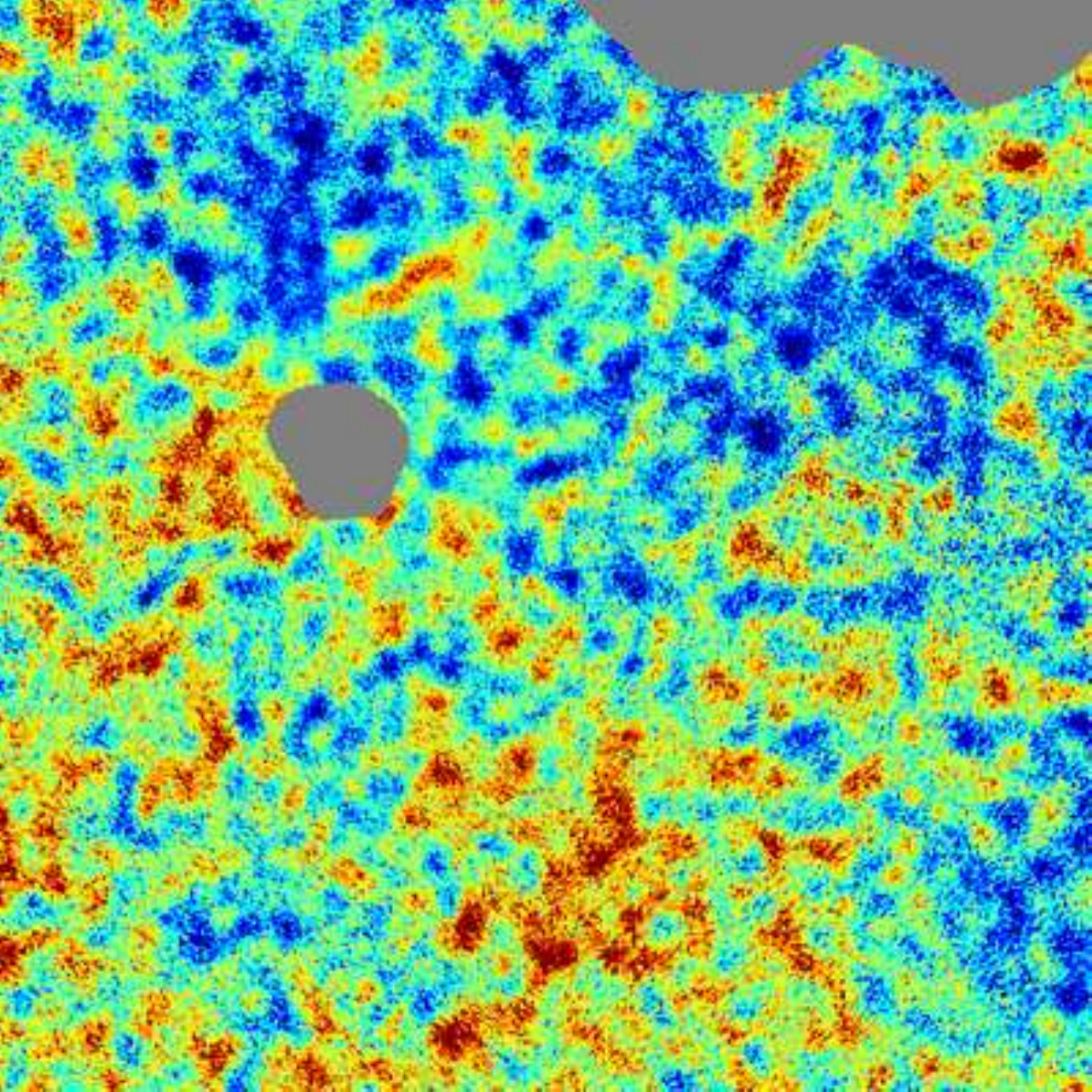}
  }
  \subfigure[WMAP-V]{
    \includegraphics[width=\widthfig]{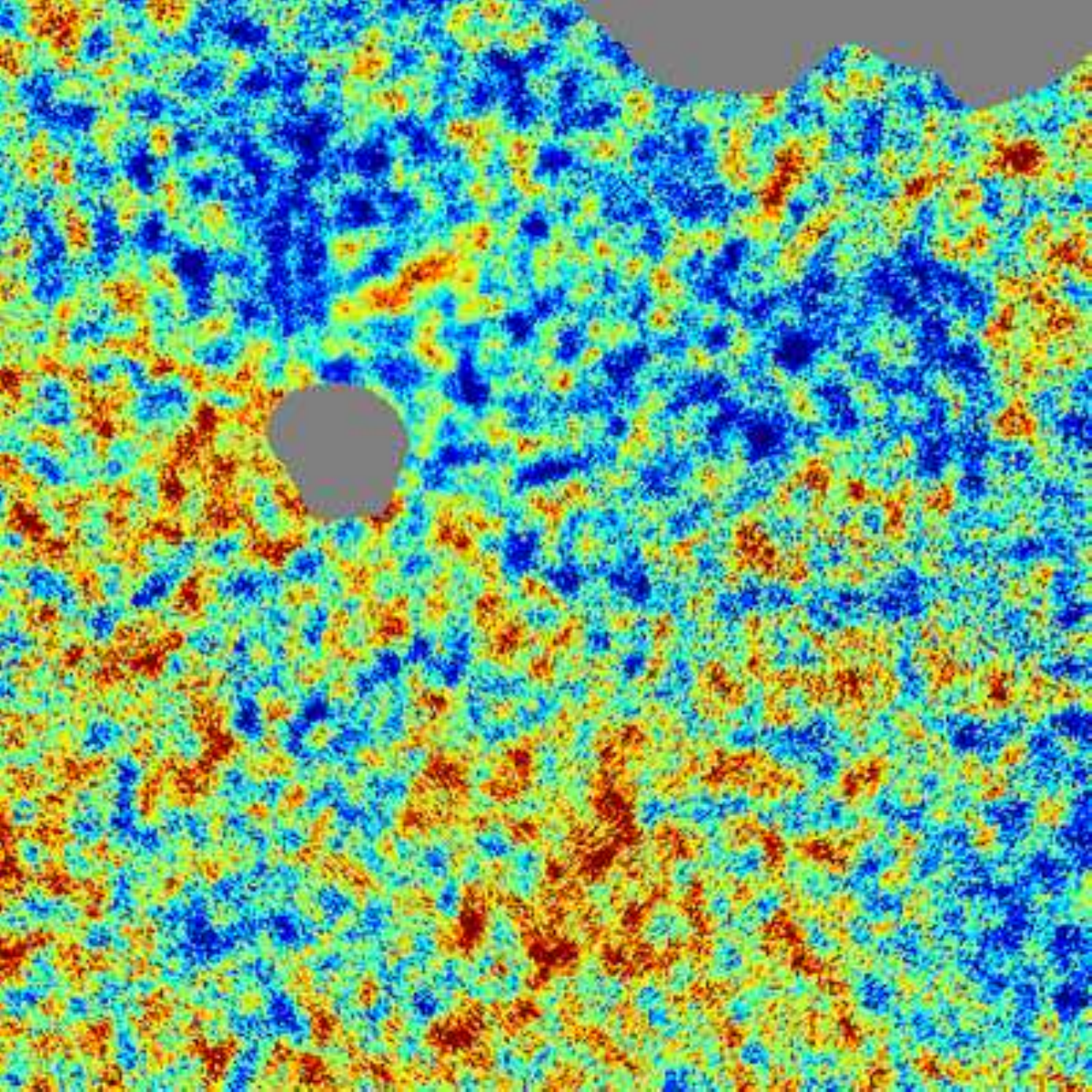}
  }
  \subfigure[WMAP-W]{
    \includegraphics[width=\widthfig]{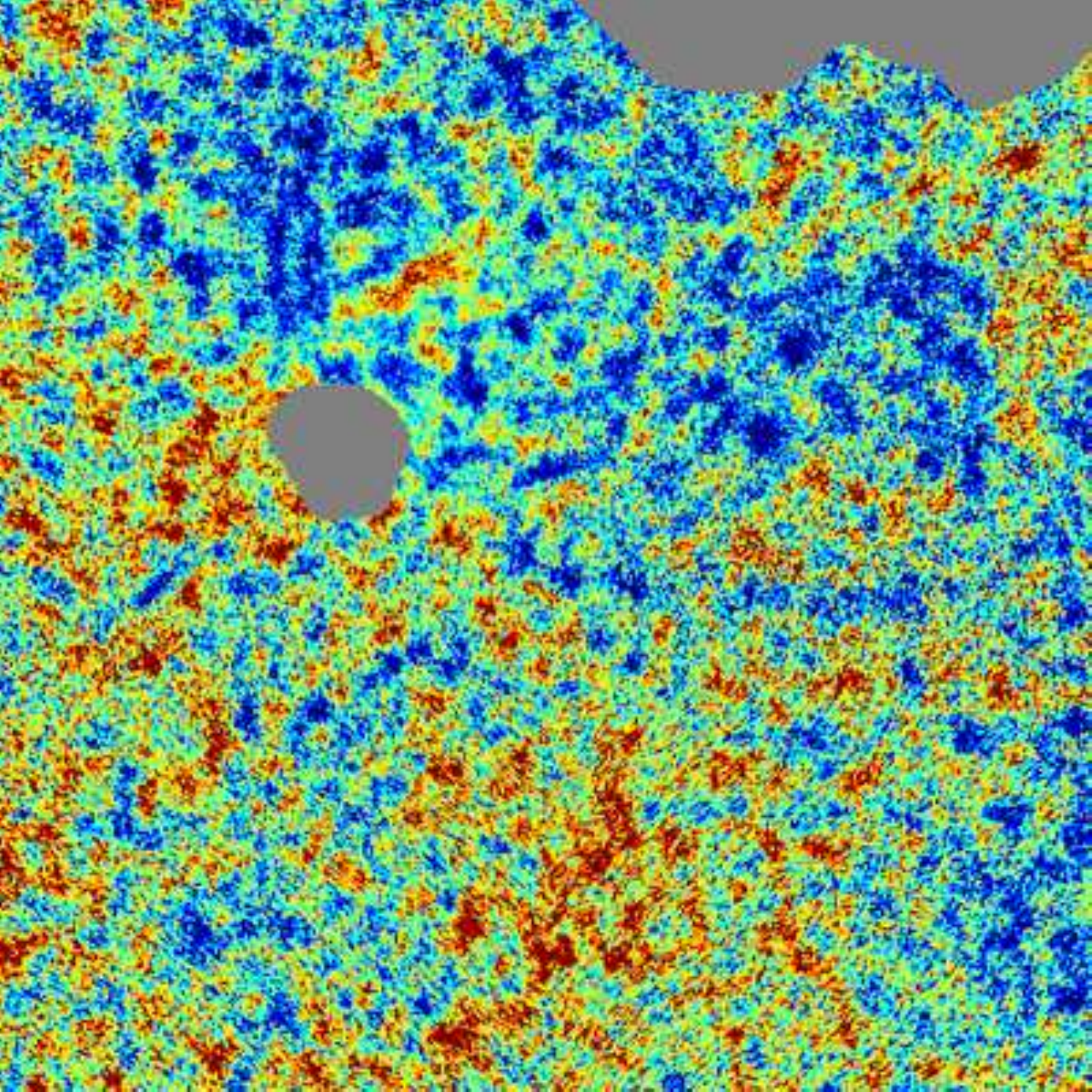}
  }\\
  \subfigure{
    \phantom{\includegraphics[width=\widthfig]{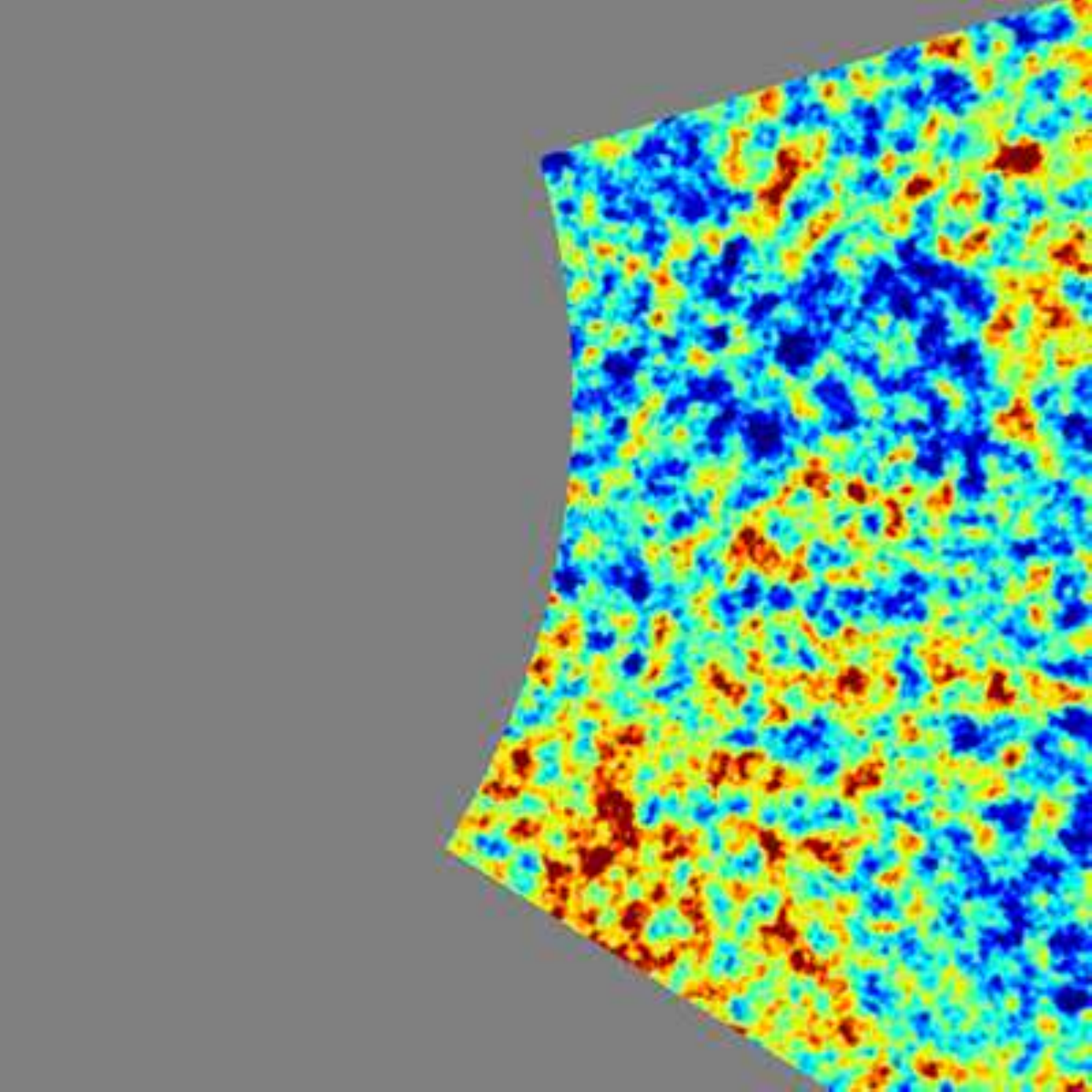}}
  }
  \addtocounter{subfigure}{-1}
  \subfigure[BOOMERanG-S]{
    \includegraphics[width=\widthfig]{simulmix_boomS.pdf}
  }
  \subfigure[BOOMERanG-D]{
    \includegraphics[width=\widthfig]{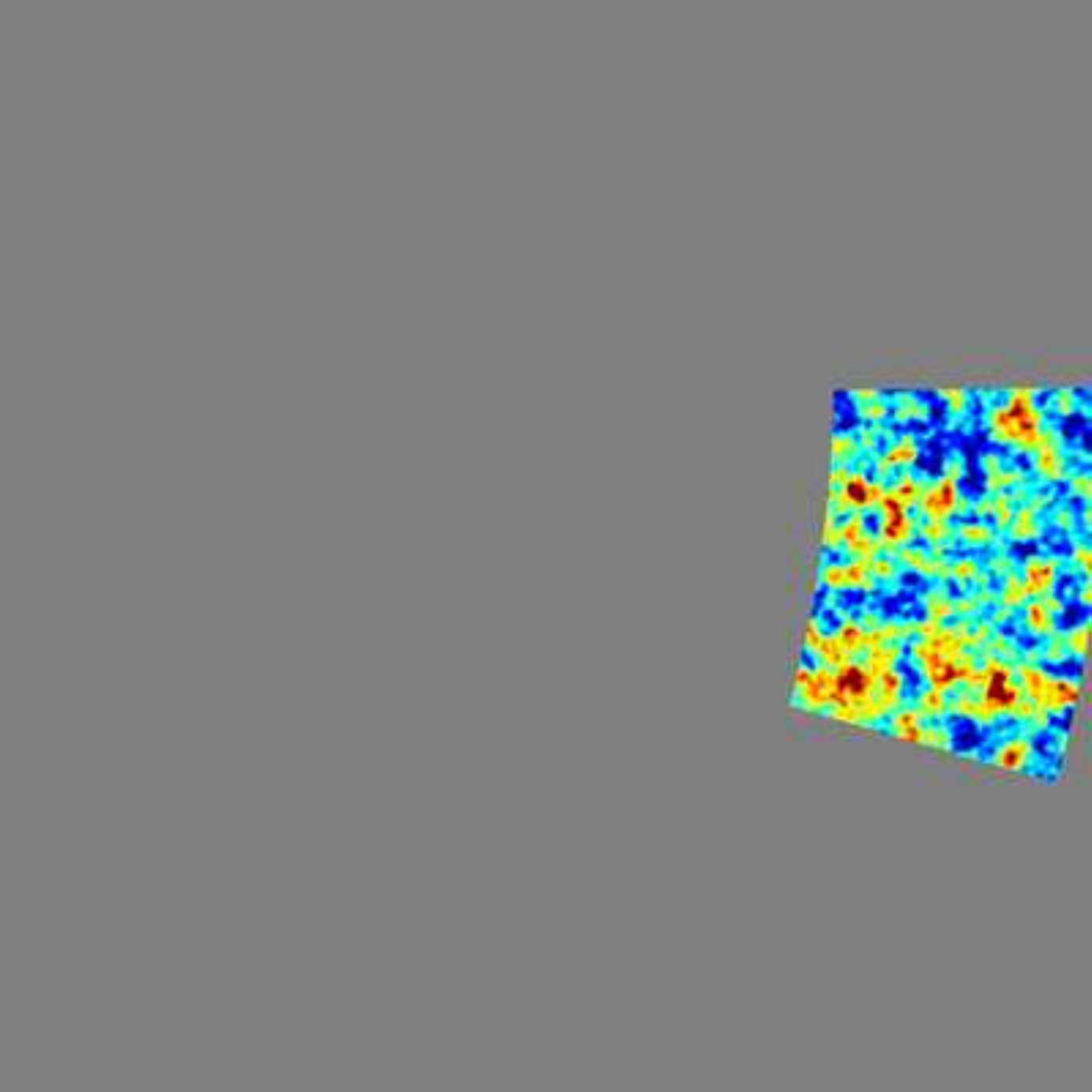}
  }
  \subfigure[ACBAR]{
    \includegraphics[width=\widthfig]{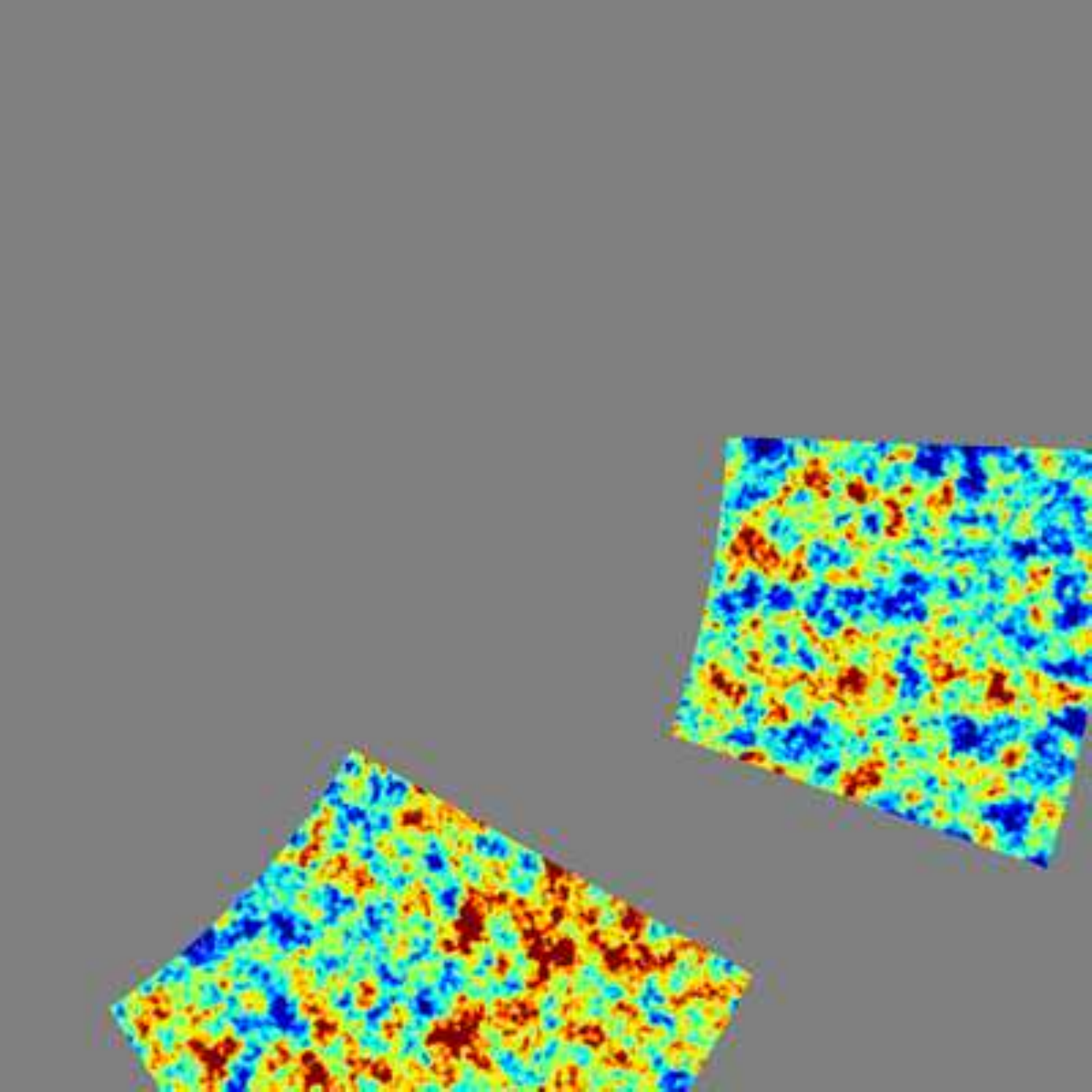}
  }
\end{minipage}
\caption{Simulated observations from model~(\ref{eq:model2}) for the 6
  experiments described in
    Table~\ref{tab:fusionparams}, in a small patch around point (-40,-90). The
    approximate size of the patch is 38$\times$38 degrees.}
  \label{fig:toy2}
\end{figure*}

\begin{figure}
  \centering
  \includegraphics[width=\linewidth]{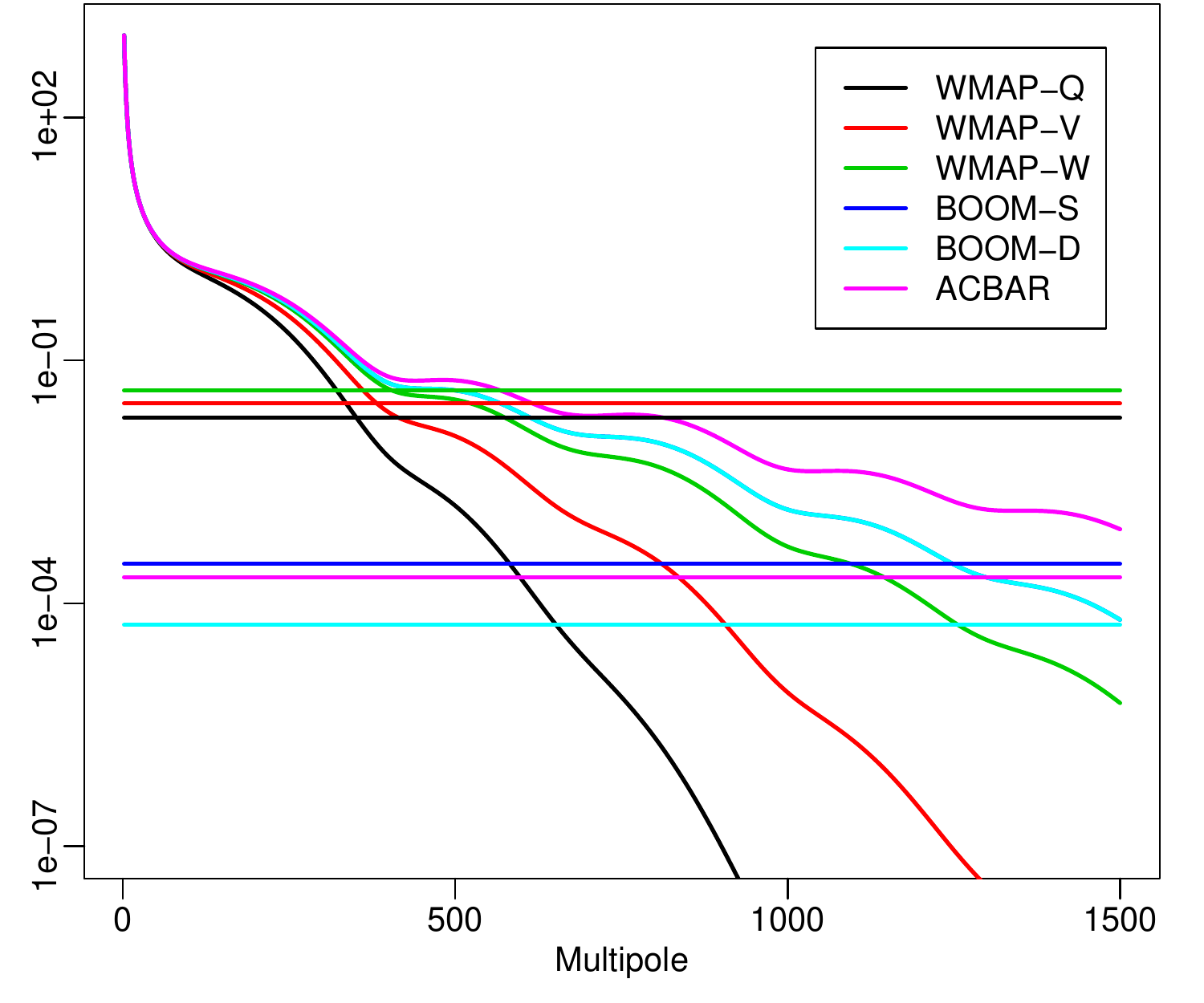}
  \caption{Spectra of the beamed CMB (with the BOOMERanG lines
    overplotted) and noise levels (horizontal lines) seen by the six
    experiments, as if they were full sky (the $\fsky$ effect is not taken into account).}
  \label{fig:spectra_exp}
\end{figure}

Key elements for this numerical experiment are illustrated in
Fig.~\ref{fig:toy2}-\ref{fig:noiselevels}-\ref{fig:spectra_exp}, where we have
displayed respectively one random outcome of each experiment according to those
simplified models, the maps of local noise levels and power spectra of the CMB
and of the experiment's noise.

\def\widthfig{0.32\linewidth}
\begin{figure*}
  \begin{minipage}{0.05\linewidth}
    \includegraphics[height=5cm,width=0.6cm]{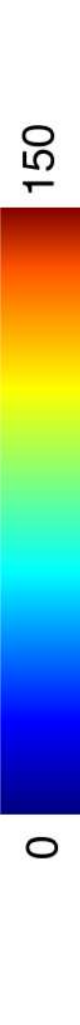}
  \end{minipage}
  \begin{minipage}{0.9\linewidth}
  \def\widthfig{5cm}
  \centering
  \subfigure[WMAP-Q]{
    \includegraphics[width=\widthfig]{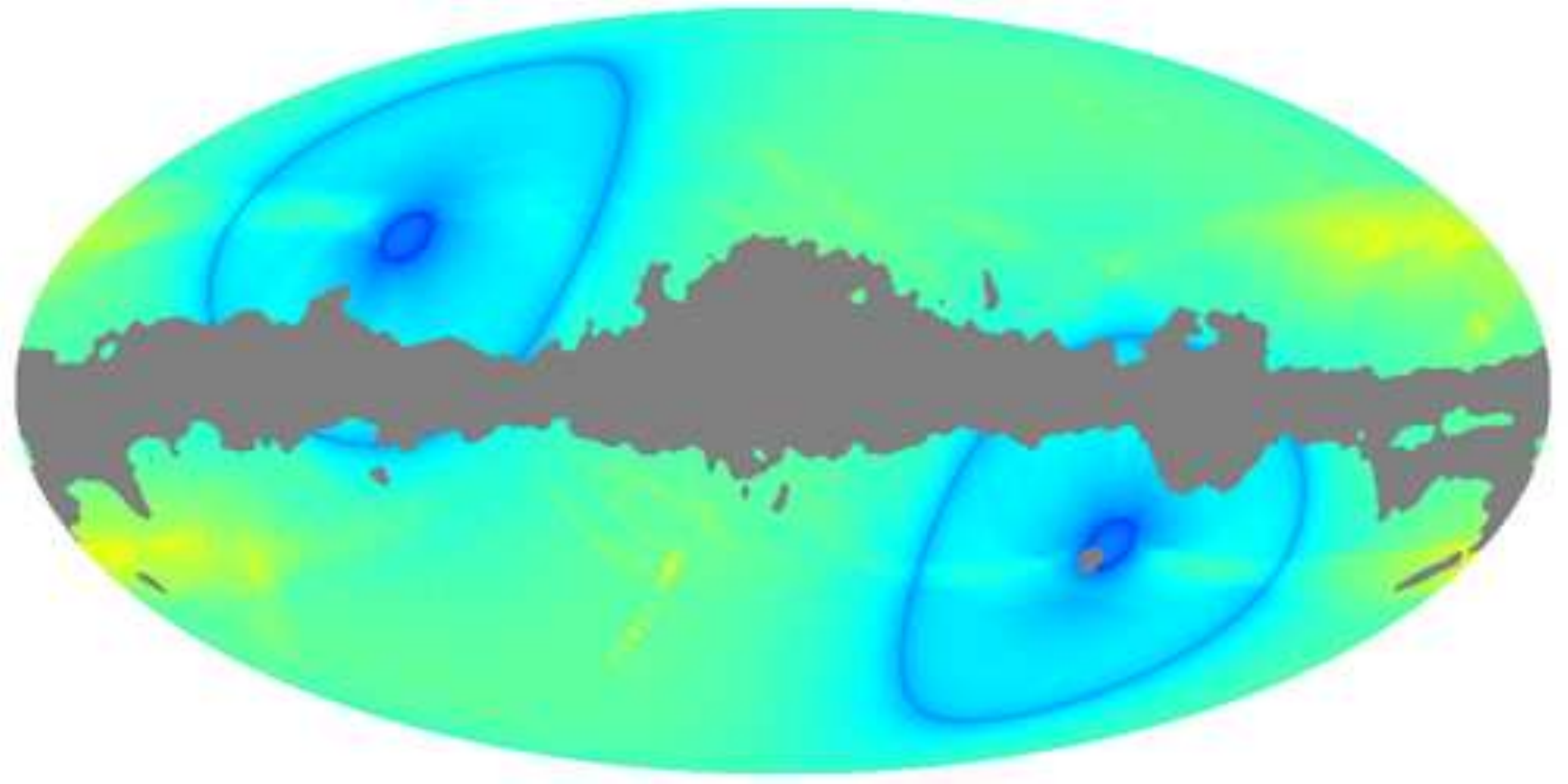}
  }
  \subfigure[WMAP-V]{
    \includegraphics[width=\widthfig]{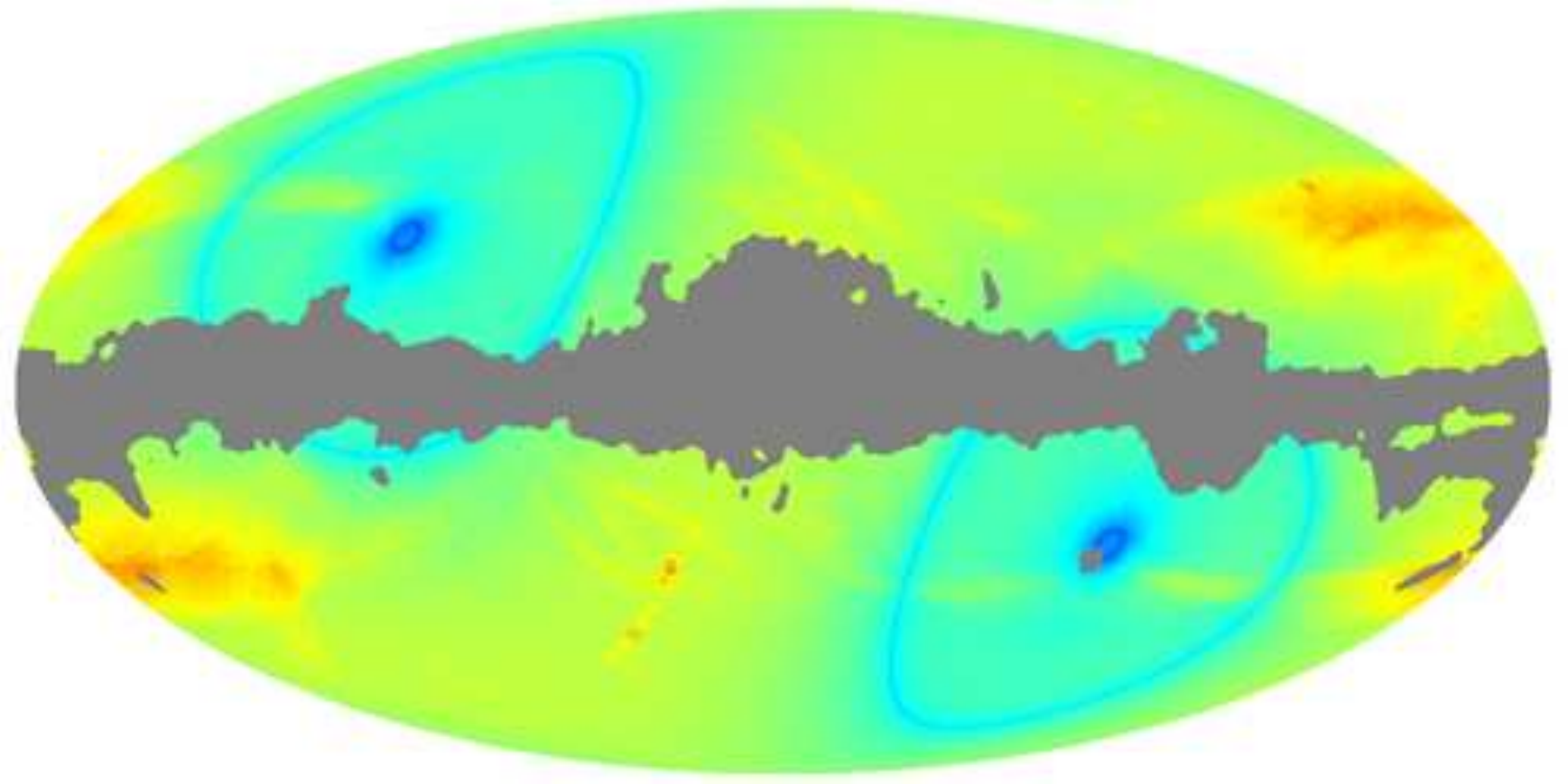}
  }
  \subfigure[WMAP-W]{
    \includegraphics[width=\widthfig]{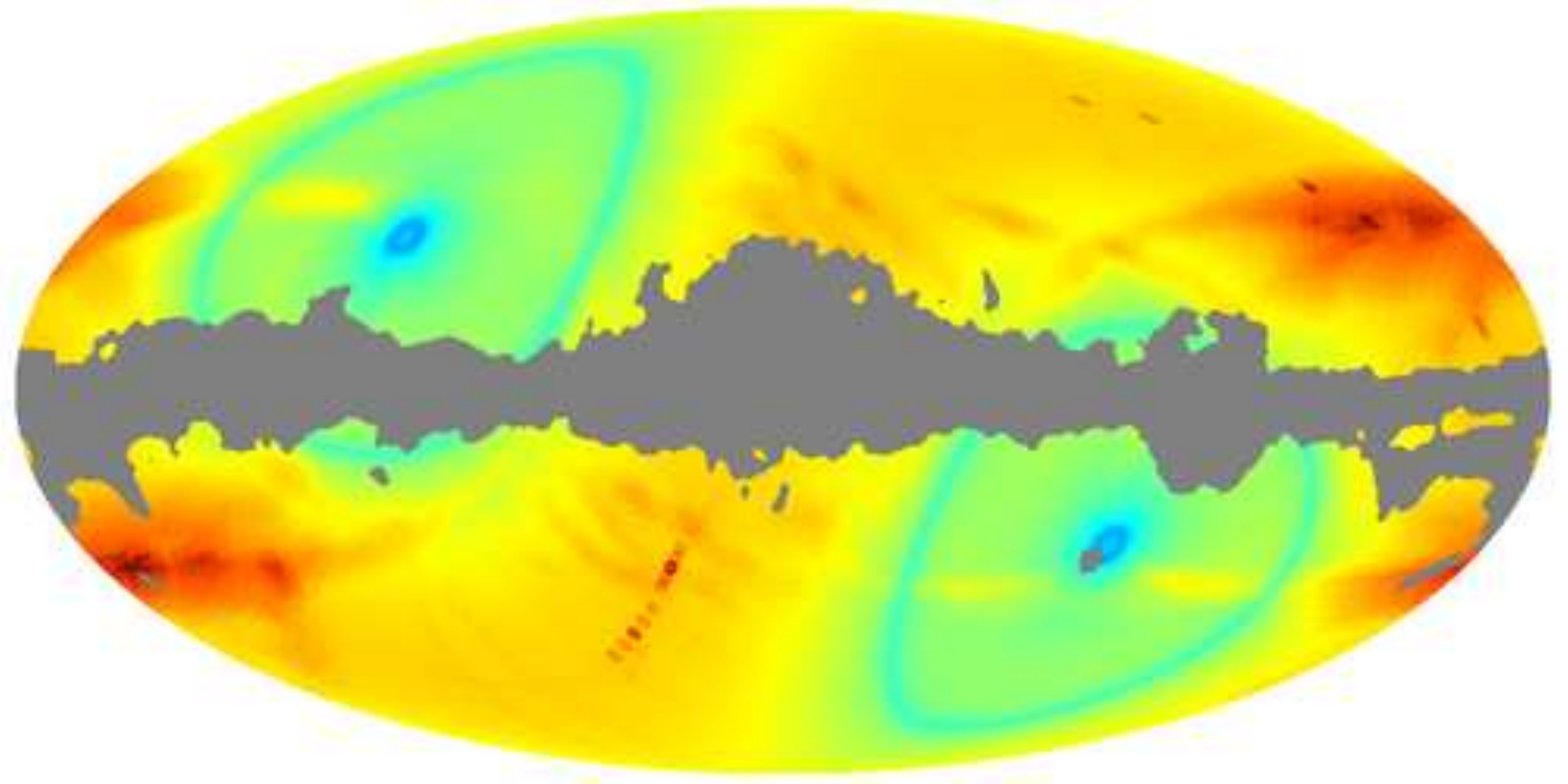}
  }
  \subfigure[BOOMERanG-S]{
    \includegraphics[width=\widthfig]{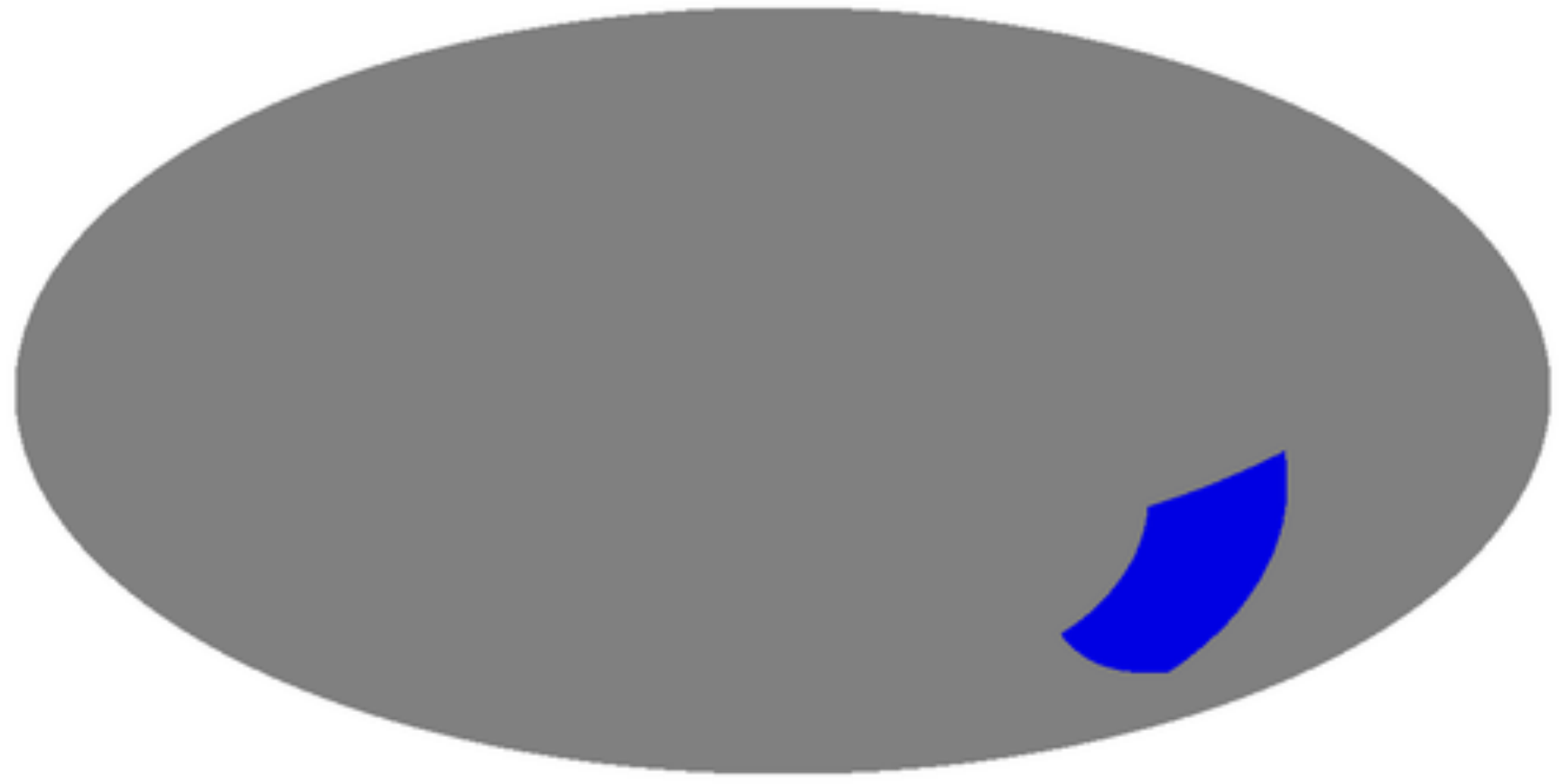}
  }
  \subfigure[BOOMERanG-D]{
    \includegraphics[width=\widthfig]{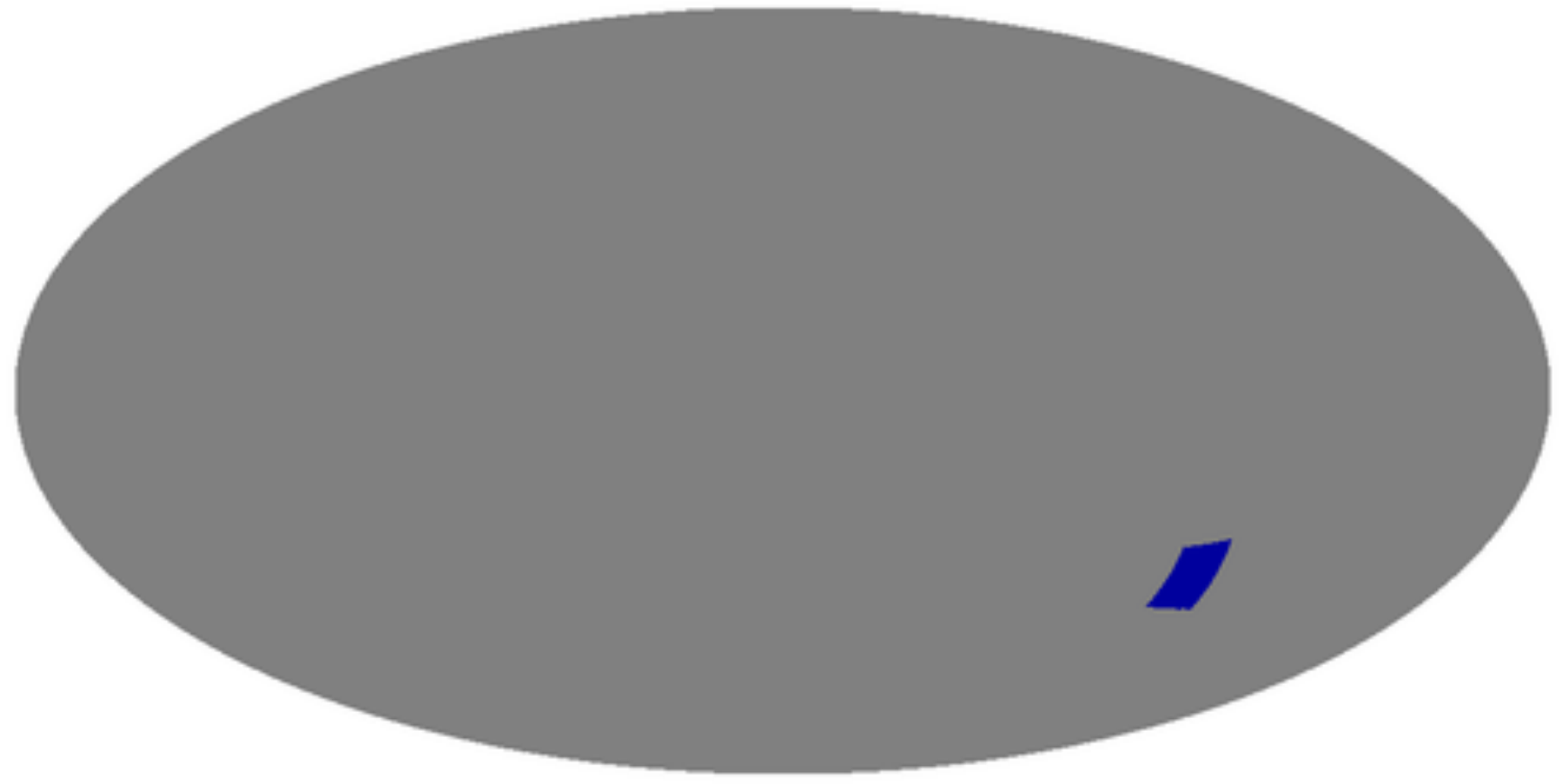}
  }
  \subfigure[ACBAR]{
    \includegraphics[width=\widthfig]{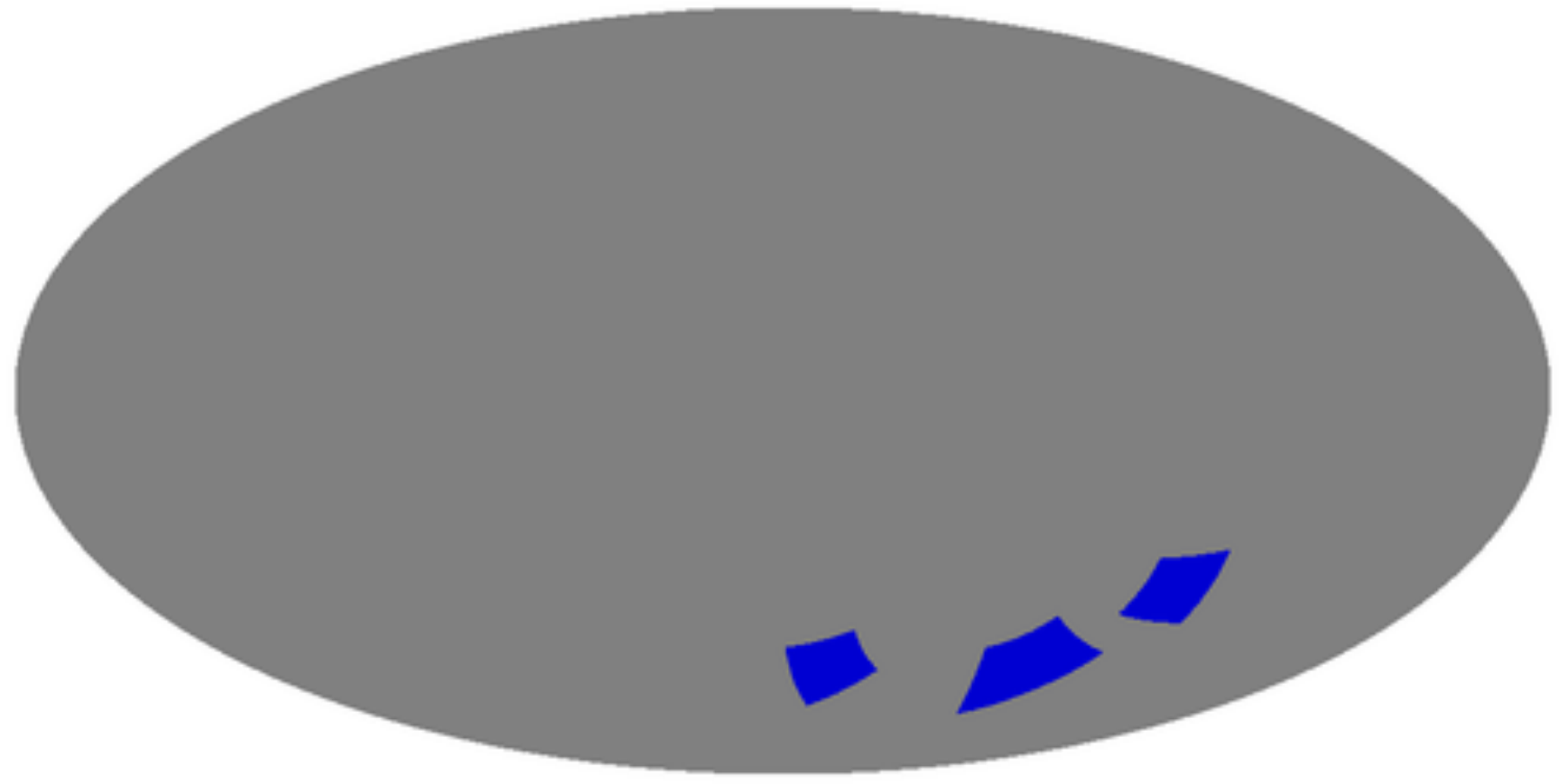}
  }
  \end{minipage}
  \caption{Coverage and local pixel noise levels of the six simple experiments described in Table~\ref{tab:fusionparams}.}
  \label{fig:noiselevels}
\end{figure*}

\begin{figure*}
  \begin{minipage}{0.05\linewidth}
    \includegraphics[height=5cm,width=0.6cm]{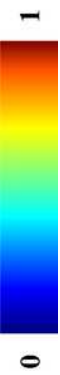}
  \end{minipage}
  \begin{minipage}{0.9\linewidth}
  \def\widthfig{5cm}
  \centering
    \includegraphics[width=\widthfig]{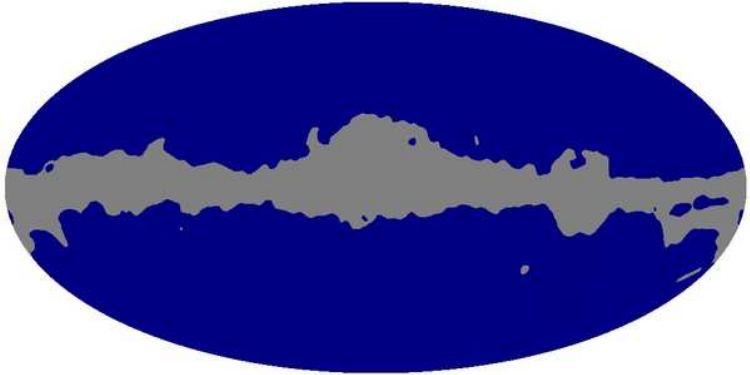}
    \includegraphics[width=\widthfig]{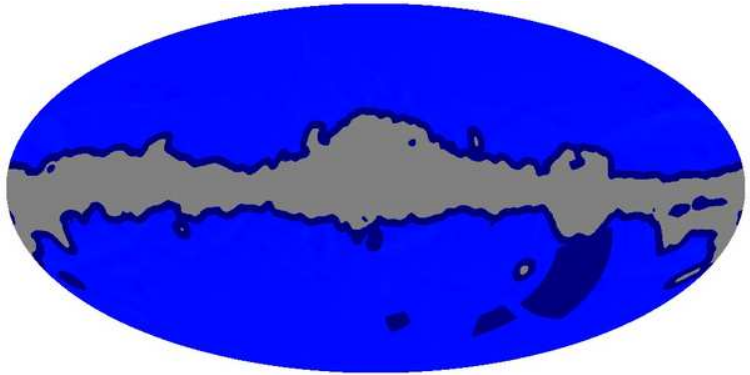}
    \includegraphics[width=\widthfig]{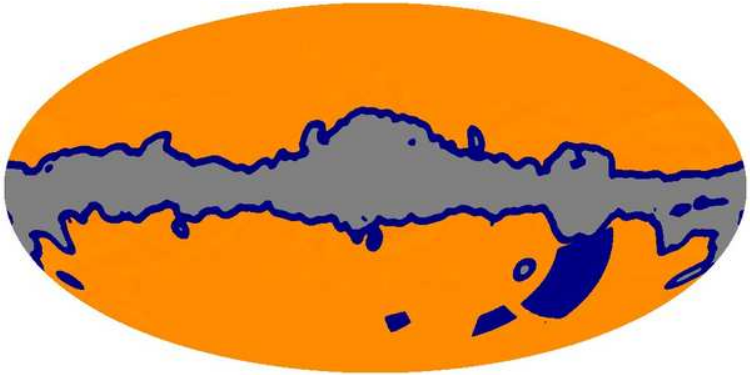}\\
    \includegraphics[width=\widthfig]{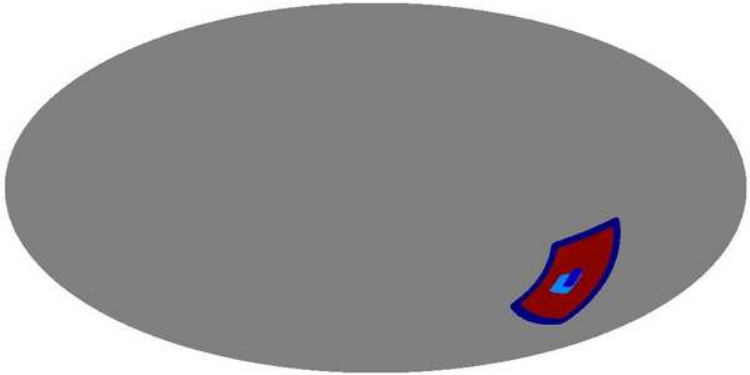}
    \includegraphics[width=\widthfig]{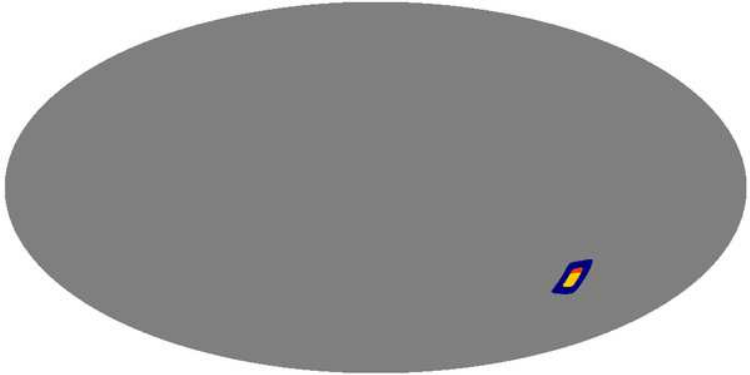}
    \includegraphics[width=\widthfig]{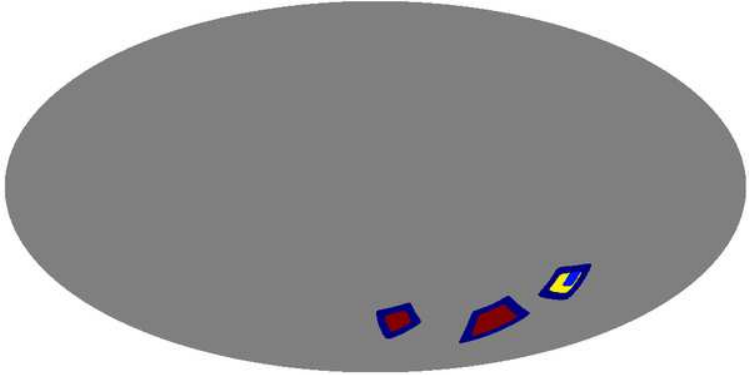}
  \end{minipage}
  \caption{Method for aggregating experiments: Weights $\omega_{k,e}^{(26)}$ for combining
    the needlet coefficients from the 26th band $(700<\ell\leq 800)$ and the
    six experiments. From left to right and top to bottom: W-MAP-Q, W-MAP-V,
    W-MAP-W, BOOMERanG-S, BOOMERang-D and ACBAR. }
  \label{fig:methodFusion1}
\end{figure*}

Fig.~\ref{fig:methodFusion1} displays the maps of the weights
$\omega_{k,e}^{(26)}$ (the 26th band is the multipole range
$700<\ell\leq800$). According to Eq.~(\ref{eq:defomegak}), all those weights
belong to [0,1] and for any fixed position, the sum of the weights over the six
experiments is equal to one. Red regions indicates needlet coefficients which
are far better observed in an experiment that in all others. Blue, light blue
and orange region are increasing but moderately low weights, showing that
outside the small patches of BOOMERanG and ACBAR, most information on band 26
is provided by the channel W of WMAP. On the patches, needlet coefficients from
W-MAP are numerically neglected in the
combination~(\ref{eq:defaggregatedcoef}).

\begin{figure*}
  \def\widthfig{\linewidth}
  \centering
  \begin{minipage}{0.49\linewidth}
    \begin{center}
    \includegraphics[width=\widthfig]{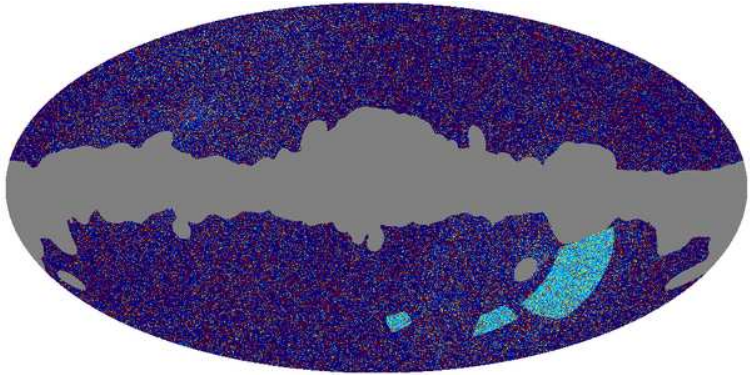}\\
    \includegraphics[width=\widthfig]{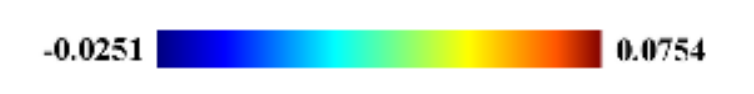}
    \end{center}
  \end{minipage}
  \begin{minipage}{0.49\linewidth}
    \includegraphics[width=\widthfig]{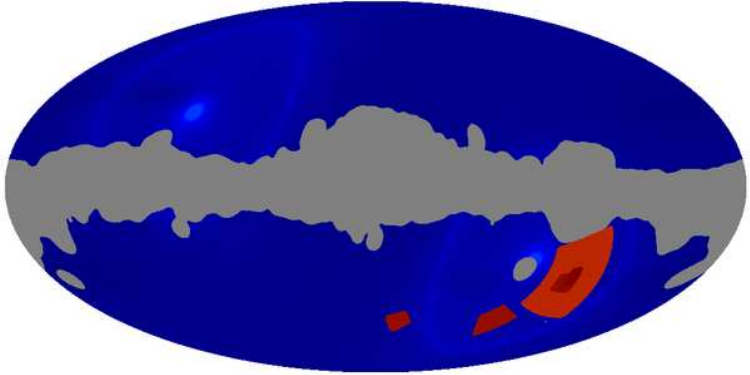} \\
    \includegraphics[width=\widthfig]{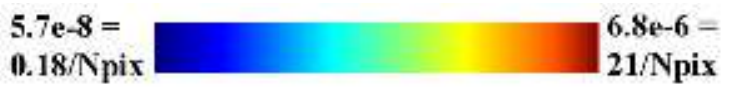}
  \end{minipage}
  \caption{Method for aggregating experiments: On the left: map of debiased
    squares of aggregated needlet coefficients, in the 26th band $(700<\ell\leq
    800)$. On the right: map of the weights $w_{k}\ji$ affected to those
    coefficients to estimate the power spectrum.}
    \label{fig:methodFusion2}
\end{figure*}

The debiased, squared, aggregated coefficients  $ \left(
    \tilde\gamma^{(26)}_k\right)^2
  -
  \left(\tilde n_k^{(26)}\right)^2$
are displayed on the left map from
Fig.~\ref{fig:methodFusion2}. All those coefficients are approximately
unbiased estimators of $C^{(26)}$. The map of weights $\tilde w_k^{(26)}$ is
displayed on the right of Figure~\ref{fig:methodFusion2}. More weight is given
to regions which are covered by lower noise experiments. The final estimate is
obtained by averaging the pixelwise multiplication of these two maps.

\begin{figure*}
  \subfigure[ACBAR]{\includegraphics[width=0.49\linewidth]{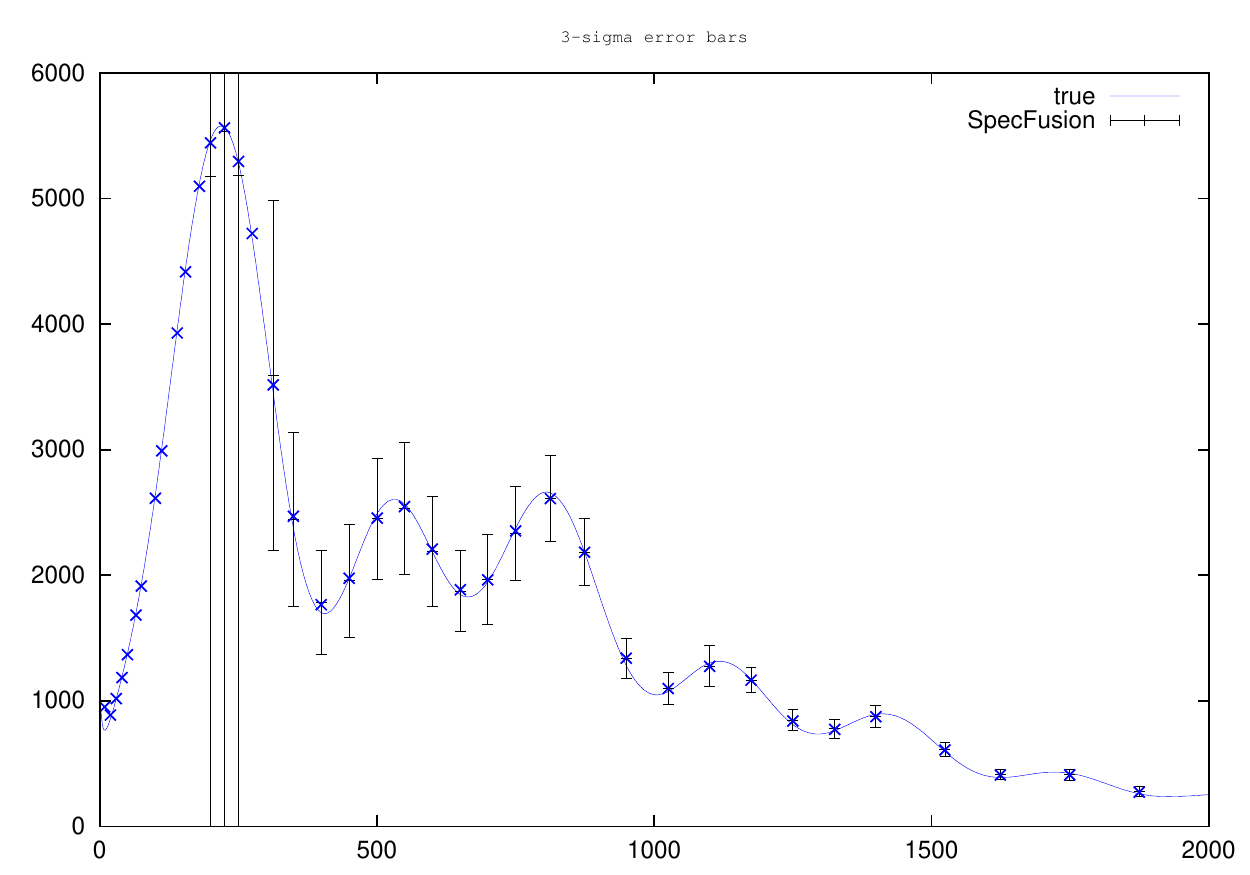}}
  \subfigure[BOOMERanG]{\includegraphics[width=0.49\linewidth]{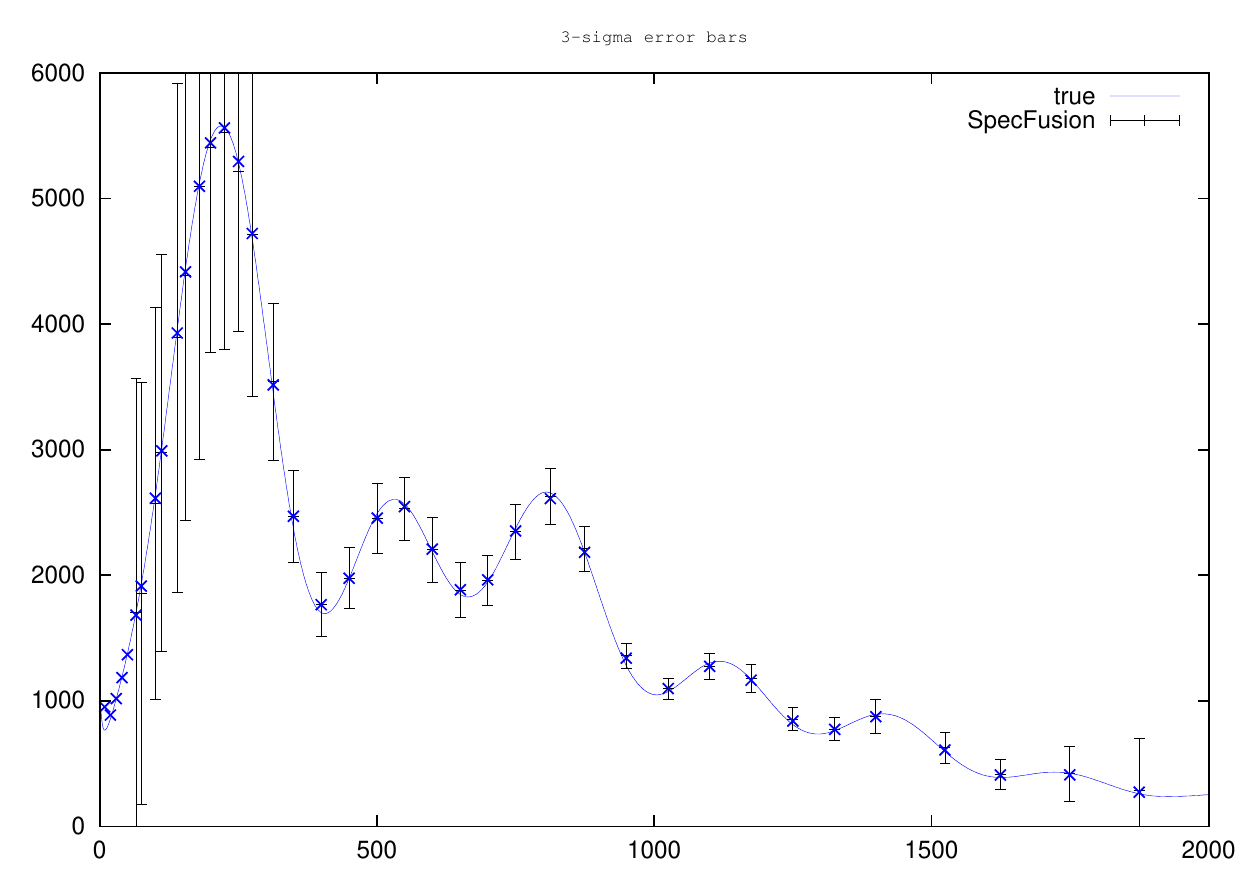}}\\
  \subfigure[WMAP]{\includegraphics[width=0.49\linewidth]{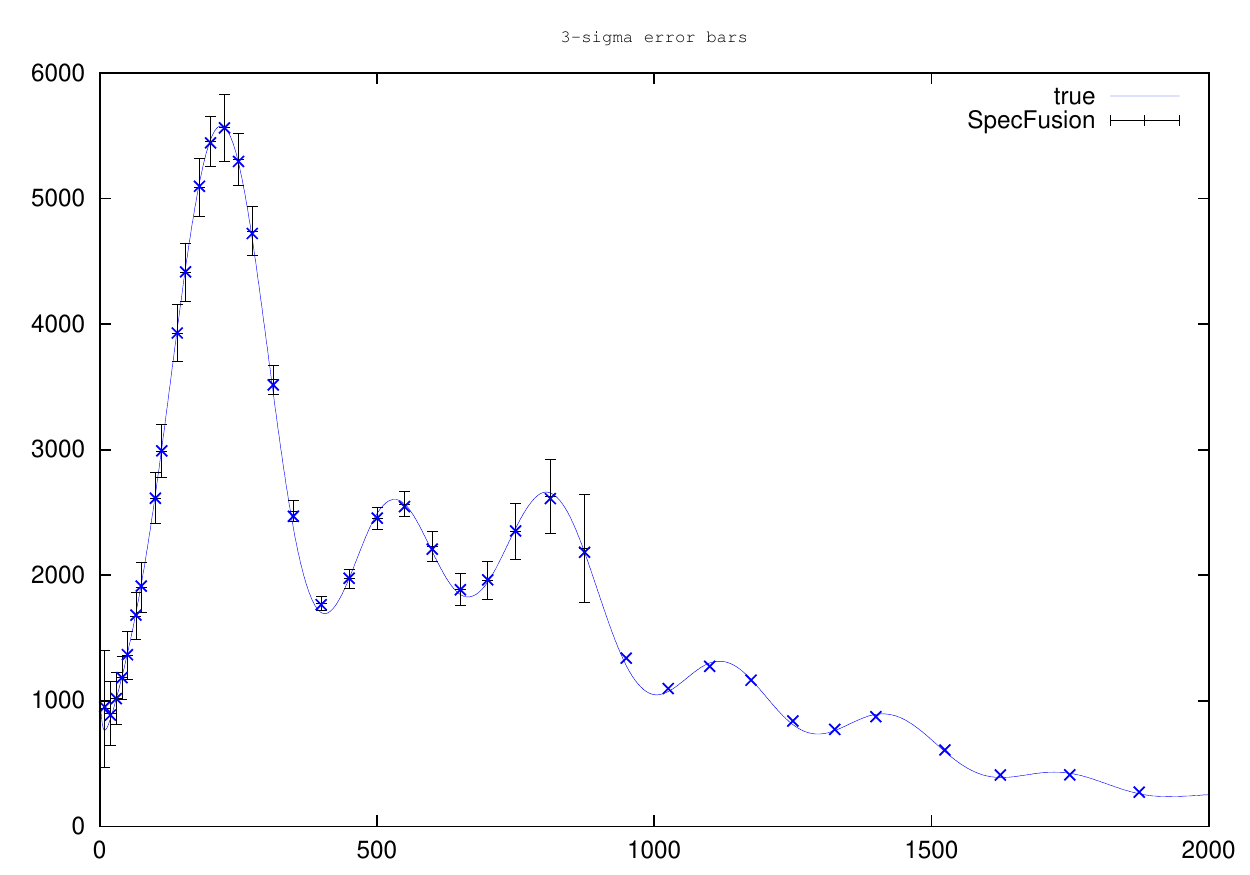}}
  \subfigure[All aggregated]{\includegraphics[width=0.49\linewidth]{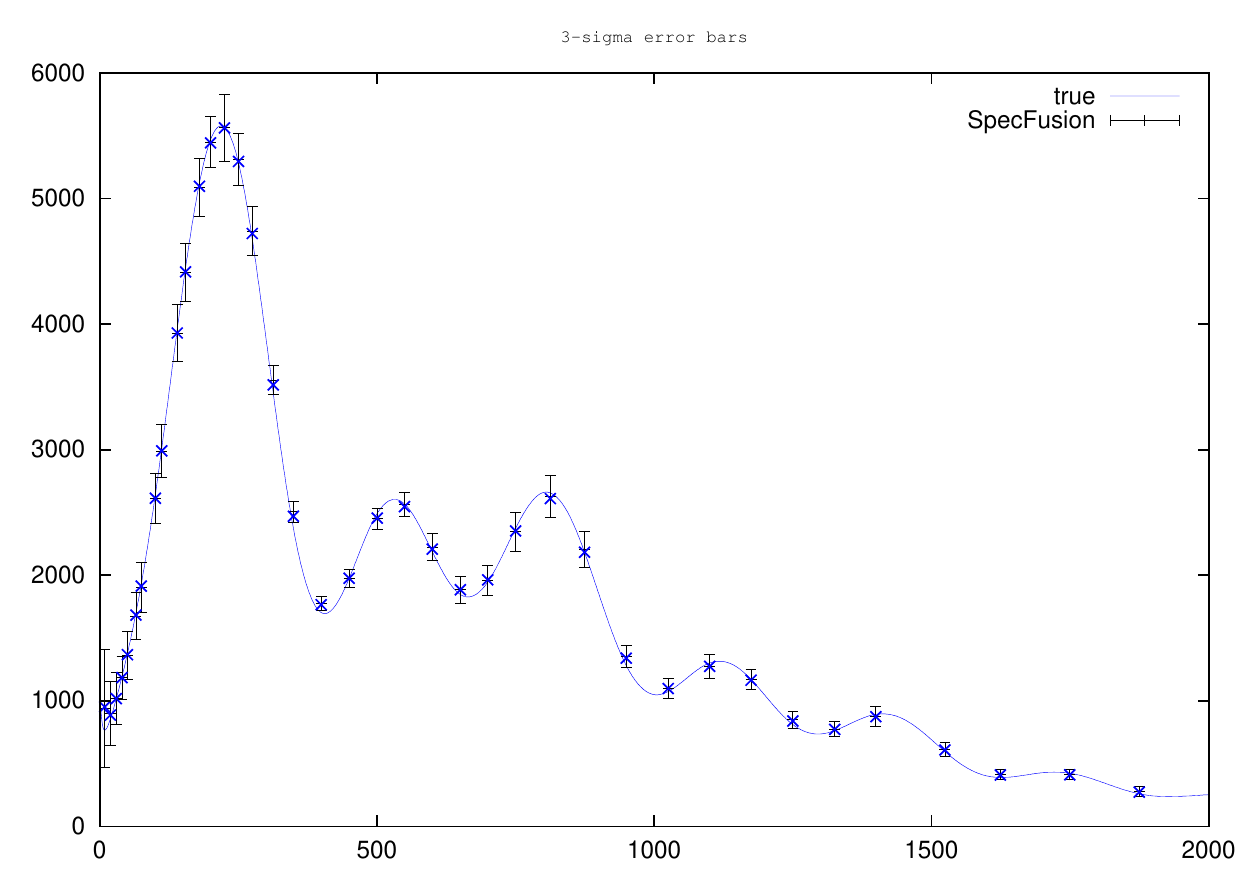}}    
  \caption{Results for the aggregated NSE. Error bars are estimated by \nmcSF{} Monte Carlo
    simulations. The ACBAR power spectrum is computed using the single map
    needlet estimator described in Section~\ref{sec:spectr-estim-single},
    whereas the BOOMERanG and W-MAP spectra are obtained using the aggregation of
    needlets coefficients from the two (BOOM-S and BOOM-D) and three (W-MAP
    Q,V,W) maps respectively. The final
    spectrum (d) is obtained by aggregating all available needlet coefficients
    from the six maps.}
  \label{fig:resultFusion}
\end{figure*}

Figure~\ref{fig:resultFusion} shows the benefit of the aggregation of different
experiments, in comparison with separate estimations.  In CMB literature, error
bars from different experiments are usually plotted on a same graph with
different colors. For easier reading, we plot the output of single experiment
NSE in separate panels (a,b and c).  Panel (d) shows the output of the
aggregated NSE, which improve the best single experiment uniformly over the
frequency range, thanks to the locally adaptive combination of informations
from all expermiments.

\begin{figure}
  \centering
  \includegraphics[width=\linewidth]{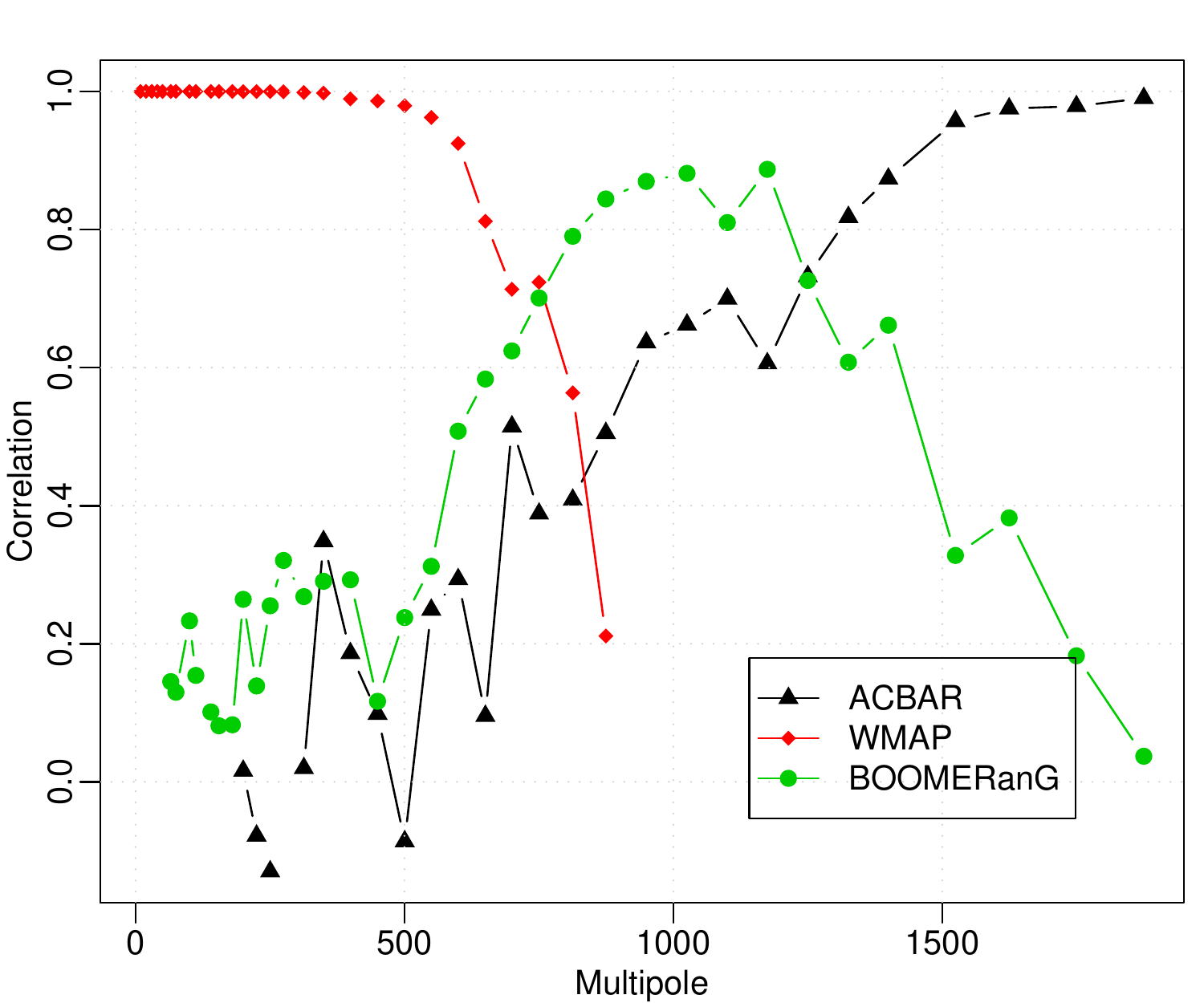}
  \caption{Correlation between the aggregated estimator and single experiments
    estimators. This provides insight on the contribution of each experiment
    into the final aggregated single spectral estimate.}
  \label{fig:corplot}
\end{figure}

Figure~\ref{fig:corplot} highlights the cross-correlation between single
experiment estimators and the final aggregated estimator. It provides a
complementary insight on the relative weight of each experiment in the spectral
domain. The W-MAP-like measures are decisive for lower bands, whereas BOOMERanG
and ACBAR ones give estimators very much correlated to the aggregated one at
higher bands. The aggregated NSE is  eventually almost identical to the
estimator obtained from ACBAR alone.


\begin{figure*}
  \centering
  \includegraphics[width=0.8\linewidth]{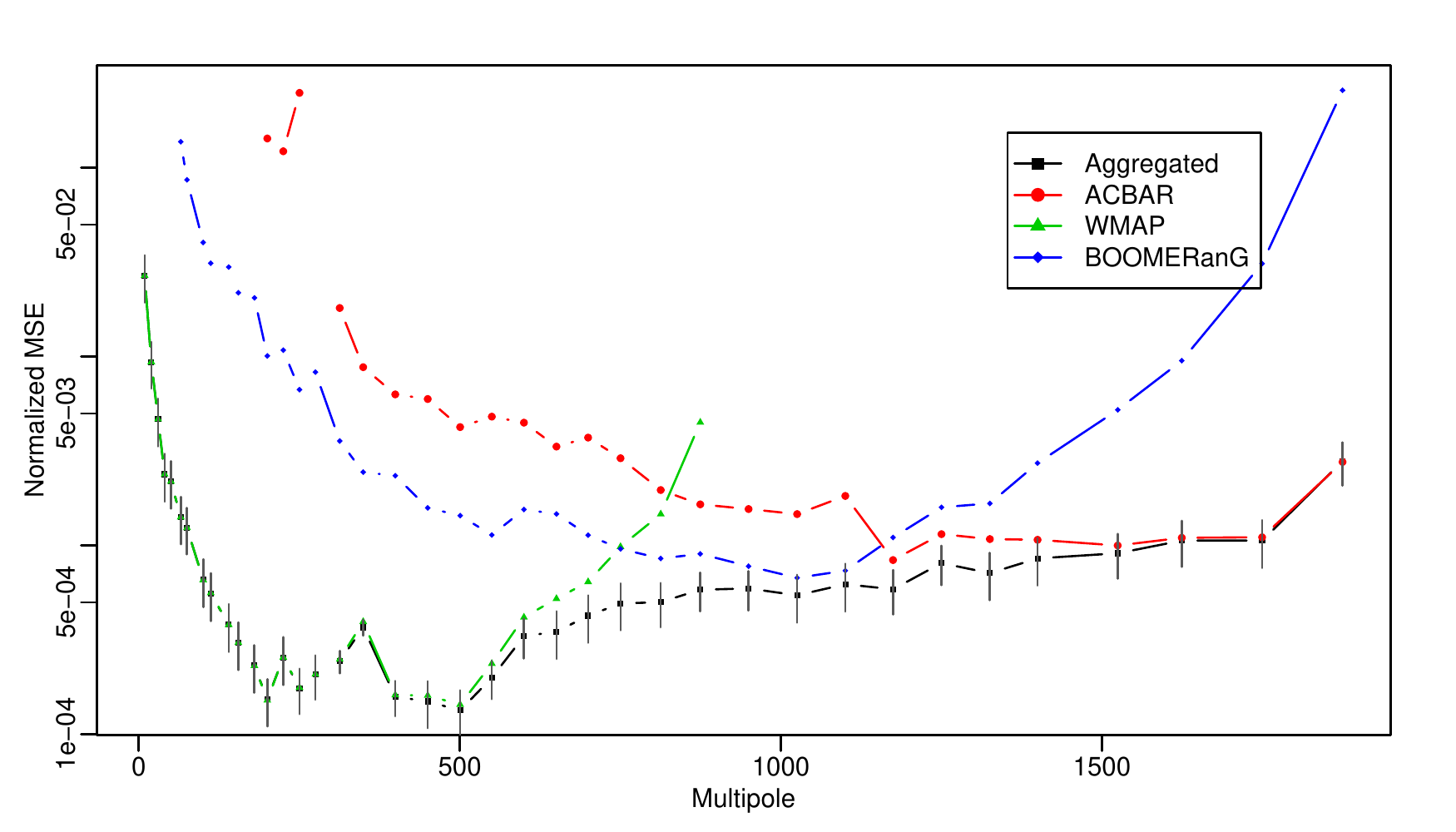}
  \caption{Mean-square error of the three single expermiment NSE estimators and
    of the aggregated NSE estimator. The 2-sigmas error bars reflect the
    imprecision in the Monte Carlo estimation of the MSE of the aggregated NSE.
  Up to those uncertainties, the aggregated estimator is uniformly better than
  the best of all experiments. The improvement is decisive in ``crossing''
  regions, where two expermiments perform comparably. The normalized MSE here
  is $\esp(\hat C\ji - C\ji)^2 / (C\ji)^2$.}
  \label{fig:mse_fusion}
\end{figure*}

\section{Discussion}
\label{sec:discussion}

\subsection*{Complexity}
According to~(\ref{eq:diagram}), the calculation of all the needlet
coefficients takes one SHT and $j_{\max}$ inverse SHT, where $j_{\max}$ is the
number of bands. The weights $w_{k}\ji$, $\tilde w_{k}\ji$ and
$\omega_{k,e}\ji$ are obtained using simple operations on maps, so that the
overall cost of the (aggregated) NSE scales as $\npix^{3/2}$ operations.  This
is comparable to the cost of the PCL methods.

\subsection*{Sensitivity to the noise knowledge}

To be unbiased, the above described estimators require a perfect knowledge of
the noise characteristics, as do Pseudo-$C_\ell$ estimators. In both cases, the
uncertainty on the noise can be tackled using cross-spectrum, that removes the
noise on the average provided that the noises from each experiment are
independent. Indeed, for any pixel $k$ far enough from the masks of experiment
$e$ and $e'$, $e\neq e'$, we have
\begin{equation*}
  \esp [\gamma_{k,e}\ji \gamma_{k,e'}\ji] = B_e\ji B_{e'}\ji C\ji.
\end{equation*}
Thus, an unbiased spectral estimator is given by
\begin{equation}\label{eq:cross_spectrum_nse}
  \widehat C\ji_{\text{cross}}
  = \sum_{k\in \mathcal{K}\ji } w_k\ji
  \sum_{e\neq e'} \left(B_e\ji B_{e'}\ji\right)^{-1}\gamma_{k,e}\ji \gamma_{k,e'}\ji
\end{equation}
where the weights $w_k\ji$ depend on a preliminary estimate of the spectrum and
a possibly imprecise estimate of the local and aggregated noise levels that
enter in the variance of $\gamma_{k,e}\ji \gamma_{k,e'}\ji$. This has not been
investigated numerically yet but we can conjecture the qualitative results of
this approach: more robustness with respect to noise misspecification but
greater error bars than the NSE with perfectly known noise levels. Moreover,
adapting the procedure described in \cite{polenta:etal:2005}, one can test for
noise misspecification, and for the correct removal of the noise by considering
the difference between the NSE $\widehat C\ji$ and cross-spectrum NSE $\widehat
C\ji_{\text{cross}}$.

\section{Conclusion}

We have presented some potentialities of the needlets on the sphere for the
angular power spectrum estimation. This tool is versatile and allows to treat
consistently the estimation from a single map or from multiple maps. 
There remains many ways of improving or modify the method described in
Section~\ref{sec:needl-spectr-estim}. 

In the future, it is likely that again complementary data sets will co-exist. This is the case, in particular, for polarisation, for which Planck will measure the large scale CMB power on large scales with moderate sensitivity, while ground-based experiments will measure very accurately polarisation on smaller scales. Extensions to polarisation of the approach presented hers will likely be important for the best exploitation of such observations.

\acknowledgements{
  The ADAMIS team at APC has been partly supported by the Astro-Map and Cosmostat ACI grants
  of the French ministry of research, for the development of innovative CMB data analysis methods.
  The results in this paper have been derived using the
  HEALPix package \citep{Gorski+2005}.
  Our pipeline is mostly implemented in octave (www.octave.org).
}

\onecolumngrid
\appendix

\section{Pseudo-$C_\ell$ estimators}
\label{sec:pseudo-c_ell-estim}

Let $T$ be a stationary process with power spectrum $(C_\ell)_{\ell \geq 0}$,
$\mcW$ an arbitrary weight function (or mask) and
$$
\widetilde C_\ell(\mcW) = \frac{1}{2\ell +1}\sum_{m=-\ell}^\ell|\langle
Y_{\ell m},\mcW T\rangle|^2
$$
the so-called \emph{pseudo-power spectrum} of $T$ with mask $\mcW$.  The
ensemble-average of this quantity is related to the true power spectrum by the
formula
\begin{equation*}
  \esp(\widetilde C_\ell) = \mcM_{\ell \ell'}(\mcW)C_{\ell'}
\end{equation*}
where $\mcM_{\ell \ell'}(\mcW)$ is the doubly-infinite coupling matrix
associated with $\mcW$, see \cite{Peebles:1973,hivon:etal:2002}. If $U$ is a
unit variance white pixel noise, denote by $V_\ell \equiv
4\pi\sigma^2/N_{\text{pix}}$ its ``spectrum'' (see
Appendix~\ref{sec:what-noise-spectrum}).  Consider now the model $X = \mcW_1 T
+ \mcW_2 U$. Then, if $\mcM_{\ell \ell'}(\mcW_1)$ is full-rank,
\begin{equation*}
  (\mcM_{\ell \ell'}(\mcW_1))^{-1} \left\{\widetilde C_\ell(\mcW_1) - \mcM(\mcW_2)_{\ell
    \ell'}V_{\ell'} \right\}
\end{equation*}
is an unbiased estimator of $C_{\ell'}$. It is obtained by deconvolving and
debiasing the empirical spectrum. The observation model (\ref{eq:model}) with
no beam coincides with the preceding framework with $\mcW_1 = W$ and $\mcW_2 =
\sigma W$. This leads to the \emph{uniform-weights pseudo-$C_\ell$ estimator}
(PCLU). One can also divide all the observations by $\sigma^2$, yielding to a
similar scheme with $\mcW_1 = \sigma^{-2}W$ and $\mcW_2 = \sigma^{-1}W$. This
is the \emph{variance-weighted pseudo-$C_\ell$ estimator} (PCLW). Both are used
by the W-MAP collaboration \citep{Hinshaw+2006}.  The uniform weights lead to
better estimates in the high SNR regime (low $\ell$'s) whereas the flat weights
perform better at low SNR (high $\ell$'s).  \cite{efstathiou:2004} showed that
the Pseudo-$C_\ell$ estimator is statistically equivalent to the maximum
likelihood estimator asymptotically as $\ell$ goes to infinity. He also
proposed an implementation of a smooth transition between those two regimes.

\section{What ``noise spectrum'' means}
\label{sec:what-noise-spectrum}
\def\noise{\nu}

Let $\noise$ denote the noise. It is defined on pixels and supposed centered,
Gaussian, independent from pixel to pixel, and of variance $\sigma^2(\xi)$, \ie
\begin{equation*}
  \noise_k=\sigma(\xi_k)U_k,\quad k=1,\dots\npix,
\end{equation*}
with $U_1,\dots,U_{\npix}\iid\mcN(0, 1)$.  Define
$\noise\lm:=\sum_k\lambda_k\noise_kY\lm(\xi_k)$, and call them (abusively) the
``discretized'' multipole moments of the noise, which do not have any
continuous counterpart because $\noise$ is not defined on the whole
sphere. Define the corresponding discretized empirical spectrum
$\overline{N_\ell}:= \frac1{2\ell+1}\sum_m\noise\lm^2$, then
\begin{gather*}
  \esp(\noise\lm \noise_{\ell'm'}) =
  \sum_k\lambda^2_k\sigma^2(\xi_k)Y\lm(\xi_k)Y_{\ell'm'}(\xi_k)\\
  \overline{N_\ell} = \frac1{2\ell+1}\sum_{k,k'}
  \lambda_k\lambda_{k'}\sigma(\xi_k) \sigma(\xi_{k'})U_kU_{k'}
  L_\ell(\langle\xi_k,\xi_{k'}\rangle)\\
  \text{and}\quad\esp(\overline{N_\ell}) =
  \frac1{4\pi}\sum_k\lambda^2_k\sigma^2(\xi_k) =: N_\ell 
\end{gather*}
This sequence $N_\ell$ can be thought of as the pixel-noise spectrum.
Note that if $\lambda_k = \frac{4\pi}{N_{pix}}$, $k=1,\dots,\npix$, then $N_\ell =
  \frac1{N_{pix}}\int\sigma^2(\xi)\dd \xi$. If the noise is
  moreover homogeneous, $\sigma(\xi)\equiv\sigma$, then $\esp(\noise\lm
  \noise_{\ell'm'})=\frac{4\pi\sigma^2}{N_{pix}}\delta_{\ell,\ell'}\delta_{m,m'}$.

\section{Variance estimation by aggregation of experiments with independent
  heteroscedastic noise}
\label{sec:vari-estim-fusion}

Consider the model
\begin{equation*}
  Y_{k,e} = X_k + n_{k,e}Z_{k,e}
\end{equation*}
where $\mathbf X := [X_k]_{k\in[1,\npix]}$ and $\mathbf Z :=
[Z_{k,e}]_{(k,e)\in[1,\npix]\times[1,E]}$ are independent, $X_k\iid\mcN(0,C)$,
$Z_{k,e}\iid\mcN(0,1)$ and the noise standard deviations $n_{k,e}$ are known.
This corresponds to the observation of the same signal $\mathbf X$ by $E$
independent experiments, the observations being tainted by independent but
heteroscedastic errors.
Let $\mathbf Y_k := [Y_{k,e}]_{(e\in[1,E]}$ be the vector of observations at
point (or index in a general framework) $k$, and let $\mathbf Y := ([\mathbf
Y_k^T]_{k\in[1,\npix]})^T$ be the full vector of observations.
The covariance matrix of $\mathbf Y_k$ is $\bR_k := \1\1^TC + \bN_k $ where
$\bN_k := \text{diag}(n_{k,e}^2)_{e\in[1,E]}$ and $\1$ is the $E\times 1$
vector of ones. 
By independence of the $\mathbf Y_k$'s, the negative log-likelihood of $C$ given $\mathbf Y$ thus writes
\begin{eqnarray*}
  \like(C) := -2\log\left(P(\mathbf Y|C)\right)
  & = &
  -2 \sum\nolimits_k\log\left(P(\yk|C)\right)\nonumber\\
  &=&
  \sum\nolimits_k \ykt \bR_k^{-1} \yk + \log\det \bR_k.
\end{eqnarray*}
Denote
\begin{equation}
  \label{eq:nk}
  \tilde n_k := \left( \1^T\bN_k^{-1}\1 \right)^{-1/2}
  = \left( \sum\nolimits_e\left({n_{k,e}}\right)^{-2} \right)^{-1/2}.
\end{equation}
It is immediate to check the following identity which will be used below:
\begin{equation*}
  \bR_k^{-1}\1 = \frac{\bN_k^{-1}\1}{1+C\tilde n_k^2}.
\end{equation*}
Define $\widehat \bR_k := \yk\ykt$. The derivative of the negative log-likelihood writes
\begin{align*}
  \like'(C) = &\sum\nolimits_k -\ykt\bR_k^{-1}\frac{\partial \bR_k}{\partial C}\bR_k^{-1}\yk
  + \tr\left(\bR_k^{-1}\frac{\partial \bR_k}{\partial C}\right)\\
  =&
  \sum\nolimits_k \tr\left(-\ykt\bR_k^{-1}\1\1^T\bR_k^{-1}\yk\right)
  + \tr\left(\bR_k^{-1}\1\1^T\right)\\
  =&
  \sum\nolimits_k \1^T\bR_k^{-1}\left(\bR_k-\widehat \bR_k\right)\bR_k^{-1}\1\\
  =&
  \sum\nolimits_k\frac
  {\1^T\bN_k^{-1}\left(\bR_k-\widehat\bR_k\right)\bN_k^{-1}\1}
  {\left(1+C\tilde n_k^2\right)^2}\\
  =&
  C\sum\nolimits_k \frac
  {\left(\1^T\bN_k^{-1}\1\right)^2}
  {\left(1+C\tilde n_k^2\right)^2}
  -
  \sum\nolimits_k \frac
  {\left(\1^T\bN_k^{-1}\yk\right)^2-\1^T\bN_k^{-1}\1}
  {\left(1+C\tilde n_k^2\right)^2}\\
  =&
  C\sum\nolimits_k \left(C + \tilde n_k^2\right)^{-2}
  -
  \sum\nolimits_k \frac
  {\left(\tilde n_k^2\1^T\bN_k^{-1}\yk\right)^2 - \tilde n_k^2}
  {\left(C + \tilde n_k^2\right)^2}.
\end{align*}
It follows that the likelihood is maximized for
\begin{equation*}
  C =  \widehat C(w) 
  := \sum\nolimits_k w_k(C,\bN) \left[
    \left(\tilde n_k^2\1^T\bN_k^{-1}\yk\right)^2 - \tilde n_k^2 
  \right]
\end{equation*}
with 
\begin{equation}
  \label{eq:weightsSE}
  w_k(C,\bN) := 
  \left(C + \tilde n_k^2\right)^{-2} 
  \left[
    \sum\nolimits_i\left(C + \tilde n_i^2\right)^{-2}
  \right]^{-1}.
\end{equation}
As the optimal weights depend on $C$, this only defines implicitly the ML
estimator.  For some approximate spectrum $C^0$, the proposed explicit NSE is
given by $\widehat C(\widehat w_k)$ with $\widehat w_k = w_k(C^0,\bN)$. 

\subsection*{Particular case of a single experiment}
\label{sec:part-case-single}
In the particular case of a single experiment ($E=1$) with heteroscedastic
noise, following the model 
\begin{equation*}
  Y_{k} = X_k + n_{k}Z_{k} \; ,
\end{equation*}
the likelihood is maximized for
\begin{equation}
  \label{eq:solutionVarianceEstimation}
  C = \widehat C(w) := \sum\nolimits_k w_k(C,\bN)\left(Y_k^2-n_k^2 \right)
\end{equation}
with $ w_k(C,\bN)$ defined by Eq.~(\ref{eq:weightsSE}), and again, assuming
that $w_k$ is poorly sensitive to $C$, the NSE is $\widehat
C\left(\widehat w_k(C^0)\right)$ for some approximate spectrum $C^0$.

\bibliography{cosmobib}
\bibliographystyle{apalike}

\end{document}